\documentstyle[aps,epsf]{revtex}
\begin{document}

\title{Parallelization of a Dynamic Monte Carlo Algorithm:
a Partially Rejection-Free Conservative Approach}
\author{G. Korniss,$^1$ M. A. Novotny,$^1$ and P. A. Rikvold$^{1,2}$}
\address{$^1$Supercomputer Computations Research Institute, Florida State 
 University, \\ Tallahasee, Florida 32306-4130}
\address{$^2$Center for Materials Research and Technology and Department of 
 Physics, Florida State University, \\ Tallahasee, Florida 32306-4350}
\date{\today}
\maketitle
\begin{abstract}
We experiment with a massively parallel implementation of an algorithm for
simulating the dynamics of metastable decay in kinetic Ising models. The 
parallel scheme is directly applicable to a wide range of stochastic cellular 
automata where the discrete events (updates) are Poisson arrivals. For high 
performance, we utilize a continuous-time, asynchronous parallel version of 
the $n$-fold way rejection-free algorithm. Each processing element carries an
$l$$\times$$l$ block of spins, and we employ the fast SHMEM-library routines 
on the Cray T3E distributed-memory parallel architecture. Different processing
elements have different {\em local} simulated times. To ensure causality, the 
algorithm handles the asynchrony in a conservative fashion. Despite 
relatively low utilization and an intricate relationship between the average 
time increment and the size of the spin blocks, we find that for sufficiently 
large $l$ the algorithm outperforms its corresponding parallel Metropolis 
(non-rejection-free) counterpart. As an example application, we present
results for metastable decay in a model ferromagnetic or ferroelectric film,
observed with a probe of area smaller than the total system.
\end{abstract}


\section{Introduction}
Fast and efficient algorithms are invaluable ingredients for large-scale 
simulations in physical sciences and engineering.  
The implementation of efficient, massively parallel algorithms for Monte 
Carlo simulations is not only an interesting and challenging problem, but also 
one of the most complex ones in parallel computing. It belongs to the class 
of parallel discrete-event simulation (sometimes referred to as
distributed simulation) which has numerous applications in engineering, 
computer science, and economics, as well as in physics \cite{Fuji}. For 
example, in lattice Ising models the discrete events are spin updates, while 
in queueing networks they are job arrivals. The dynamics of these systems, 
which obviously contain a substantial 
amount of parallelism, were traditionally simulated on serial computers. 
Paradoxically, it is fairly difficult in practice to implement an efficient 
parallel algorithm to simulate these dynamics, mainly due to the fact that the 
discrete events (updates) are not synchronized by a global clock.
Here we present and analyze the performance of a variation \cite{Luba} of the 
$n$-fold way algorithm \cite{BKL,Mark} to simulate magnetization switching in 
a kinetic Ising model on a distributed-memory parallel computer.

Metastability and hysteresis are widespread phenomena in nature. Ferromagnets 
are common systems that exhibit these behaviors
\cite{ferro_magn}, but there are also numerous other examples ranging from 
ferroelectrics \cite{ferro_el} and electrochemical adsorbate layers 
\cite{electro} to liquid-crystals \cite{liquid}. An important
potential technological application of nanoscale ferromagnetic
particles and ultrathin films is high-density magnetic recording 
media, and computational experiments and modeling of these materials are 
integral parts of current research and engineering.
 
Simulating metastable decay often involves long characteristic time scales
(the lifetime of the metastable phase), and several sophisticated algorithms 
have been developed for serial computers \cite{BKL,Mark,group1} to tackle this 
problem. Common testbeds for these algorithms are kinetic Ising 
ferromagnets below their critical temperature $T_c$, which exhibit slow 
metastable decay after reversal of the external magnetic field 
\cite{switch}.
These models are appropriate for the study of highly anisotropic single-domain 
nanoparticles and thin films \cite{group2}.
 
There are powerful techniques, such as multi-spin coding \cite{multispin} and
cluster algorithms \cite{cluster,cluster_rev}, to simulate the 
{\em equilibrium} properties of the Ising model, but these methods completely 
change the original microscopic {\em dynamics}.
Kinetic Ising models, either with integer-time updates or with
Glauber's continuous-time interpretation \cite{Binder_MC}, 
were believed to be inherently serial, 
i.e., the corresponding algorithm could attempt to update only one spin 
at a time. Providing a counter example to that belief, Lubachevsky presented 
an efficient conservative approach for
parallel simulation of these systems \cite{Luba} {\em without} changing the 
dynamics of the underlying model. Applications of his scheme also include 
modeling of wireless cellular communication networks \cite{GLNW} and 
ballistic particle deposition \cite{LPR}. Also, he {\em proposed} a way to 
incorporate the $n$-fold way algorithm, possibly further contributing to 
speedup. Here, we implement this algorithm on the isotropic, two-dimensional 
Ising model, and we systematically compare its performance to the parallel 
Metropolis algorithm. To our knowledge, this is the first actual 
implementation of the parallel $n$-fold way scheme. 
More importantly, detailed comparison between the two parallel
schemes has not been given before. As we show in this paper, there is an 
interesting competition between their performances, and detailed analysis
is essential to decide which one to apply in different parameter regimes.

This paper is organized as follows. In Section II we define the model and 
summarize the standard Metropolis and rejection-free serial algorithms. 
In Section III we outline the basic conservative approach for parallel
discrete-event simulation, applied to Ising spins, and 
describe the parallel Metropolis and $n$-fold way algorithms. 
In Section IV we give some details of the implementation  and analyze its 
performance. In particular, we compare it to that of the parallel Metropolis 
update scheme. In Section V we give some examples of ongoing and future 
applications to model magnetic and ferroelectric systems, and in Section VI we 
conclude with a brief summary. 
 
\section{The Model}

We simulate magnetization switching in the isotropic Ising model on a 
$L$$\times$$L$ square lattice with periodic boundary conditions, which has the 
following Hamiltonian:
\begin{equation}
{\cal H}=-J \sum_{\langle i j \rangle} s_i s_j -H \sum_{i=1}^{L^2} s_i \;.
\label{Ising}
\end{equation}
Here $J>0$ is the ferromagnetic nearest-neighbor spin-spin interaction,
$H$ is the external field, the sum $\sum_{\langle i j \rangle}$ runs over
all nearest-neighbor bonds, and $\sum_{i}$ runs over all $L^2$ lattice sites. 
To study metastable decay, all spins are
initialized in the $+1$ state, and we apply a negative magnetic field 
(i.e., a sudden field reversal) at constant $T<T_c$. Here $T_c$ is the 
critical temperature of the zero-field Ising model, below which the system 
spontaneously orders. 
For $T$$<$$T_c$ the decay of the metastable phase proceeds through 
nucleation and growth of one or more compact droplets of the stable phase. 
For fixed $T$ and $H$ there exists a system size, approximately the
typical droplet separation $R_{o}(T,|H|)$, such that
for $L$$>$$R_{o}$, many droplets form and contribute to the decay of the 
metastable phase. This is called the multidroplet regime \cite{switch}. For
the simulations presented in this paper the parameters are chosen such that the
system is in this regime. In particular, $R_{o}$$\approx$$275$ at 
$T$$=$$0.6T_c$ and $|H|/J$$=$$0.2222$, and  $R_{o}$$\approx$$12$ at 
$T$$=$$0.8T_c$ and $|H|/J$$=$$0.4127$ are the largest and smallest values of 
$R_{o}$ for the temperatures and fields used in the following sections.

\subsection{The serial Metropolis algorithm}
In the standard serial Monte Carlo (MC) algorithm \cite{Binder_MC}
where time is a discrete variable taking on integer values, we choose 
the single-spin-flip Metropolis rates, according to which at every MC step 
(MCS) a spin is picked randomly on the lattice and flipped with the probability
\begin{equation}
p=\min\{1,\exp(-\Delta {\cal H}/kT)\} \;,
\label{metropolis}
\end{equation}
where $\Delta {\cal H}$ is the energy change which would result from the flip.

It is important to note that in this standard algorithm, the choice of fixed 
integer time increments (i.e., $\Delta t\equiv 1$ MCS between two successive 
spin-flip trials) is a convenience rather than a necessity: 
the underlying {\em dynamics} of the real physical
system corresponds to a continuous time evolution, in which spin flips
are Poisson arrivals. Thus, the time increment between two successive events
is an exponentially distributed random variable. To exactly mimic the Poisson 
arrivals one should generate random time increments by
\begin{equation}
\Delta t=-\ln(r) \;
\end{equation}
in units of MCS, where $r$ is uniformly distributed in $(0,1)$.
Clearly, this is computationally more expensive than the simple integer-time
update and usually yields identical results when averaged over many 
independent runs. This extra cost, however, can pay off when mapping the 
system onto a parallel computer, as we shall see in Sections III and IV.
 
An important quantity of interest is the average lifetime 
$\langle \tau \rangle$ of the system, which is the time needed to exit the 
metastable phase. A possible way to estimate this quantity is to keep track of
the time series of the magnetization
\begin{equation}
m=\frac{1}{L^2} \sum_{i=1}^{L^2} s_i \;,
\label{magn}
\end{equation}
which approximately equals $-1$ in the equilibrium phase.
An average of its  first-passage time to zero (i.e., \mbox{$m(t=\tau)=0$})
over many independent escapes from the metastable state will then yield  
$\langle \tau \rangle$.

The weakness of the standard Metropolis algorithm (with both integer and 
continuous time) is the low acceptance rate of spin-flip trials at low 
temperature and low field, which is a result of the small flipping probability 
$p$ in Eq.\ (\ref{metropolis}). For example, at $T$$=$$0.7T_{c}$ and 
$|H|/J$$=$$0.2857$ the fraction of successful spin-flip attempts is only about
$5\%$. One may overcome this ``waste of trials'' by using a rejection-free 
update scheme as summarized in the following subsection.

\subsection{The serial $n$-fold way algorithm} 
In the the $n$-fold way update scheme \cite{BKL}, a spin
flip is always performed, and the simulated time is incremented appropriately.
To implement the scheme, 
one must introduce the notion of spin classes which carry the
state of the spin itself and its neighbors. In the above model
there are ten such classes, characterized by the number
of spins in class $i$, $n_i$, and the flipping probability, $p_i$,  which is
the same for all spins in a class. Since the classes are disjoint, 
\mbox{$\sum_{i=1}^{10} n_i =L^2$}. 
When performing an update, a class is first chosen according to the
relative weights $\{n_i p_i\}_{i=1}^{10}$, then one of the spins in the class
is picked with equal probability, $1/n_i$.  Once the information on the 
classes has been updated, in particular the $n_i$'s, the time of the next 
update is determined. As in the standard (non-rejection-free) update scheme, 
the $n$-fold way algorithm  can be performed in either integer 
or continuous time \cite{Mark}. 
In both cases the time increment is a {\em random} variable. For the 
integer-time case, it is a geometrically distributed random number,
\begin{equation}
\Delta t =  \mbox{INT} \left[ \frac{ \ln(r)}{ \ln(1-\Gamma)} \right] +1\;,
\label{int_time}
\end{equation}
while for continuous time it is exponentially distributed,
\begin{equation}
\Delta t = - \frac{\ln(r)}{\Gamma} \;.
\label{cont_time}
\end{equation}
Here, 
\begin{equation}
\Gamma=\frac{\sum_{i=1}^{10} n_i p_i}{L^{2}} \;
\label{ave_incr}
\end{equation}
is the inverse of the average time needed to exit the
configuration specified by $\{n_i\}$, and $r$ is a uniformly distributed 
random number on $(0,1)$ \cite{BKL,Mark}. For both cases, $\Delta t$ is given 
in units of MCS. At low temperatures the $p_{i}$'s
of the dominant classes (the ones with high populations) can be very small, 
resulting in large typical time increments. For example, at $T$$=$$0.7T_{c}$ 
and $|H|/J$$=$$0.2857$ the mean time increment is approximately $20$ MCS.  
This is how the algorithm ``bypasses'' a large number of
unsuccessful flip attempts and can increase the efficiency by several
orders of magnitude \cite{Mark}.

Finally, it is important to note that both the standard Metropolis and the 
$n$-fold way algorithms yield identical physical results. They are simply two 
different implementations for simulating the same dynamics.

\section{Parallelization}

\subsection{The parallel Metropolis algorithm}
\begin{figure}[t]
\begin{center}
\epsfxsize=12.0cm\epsfysize=12.0cm\epsfbox{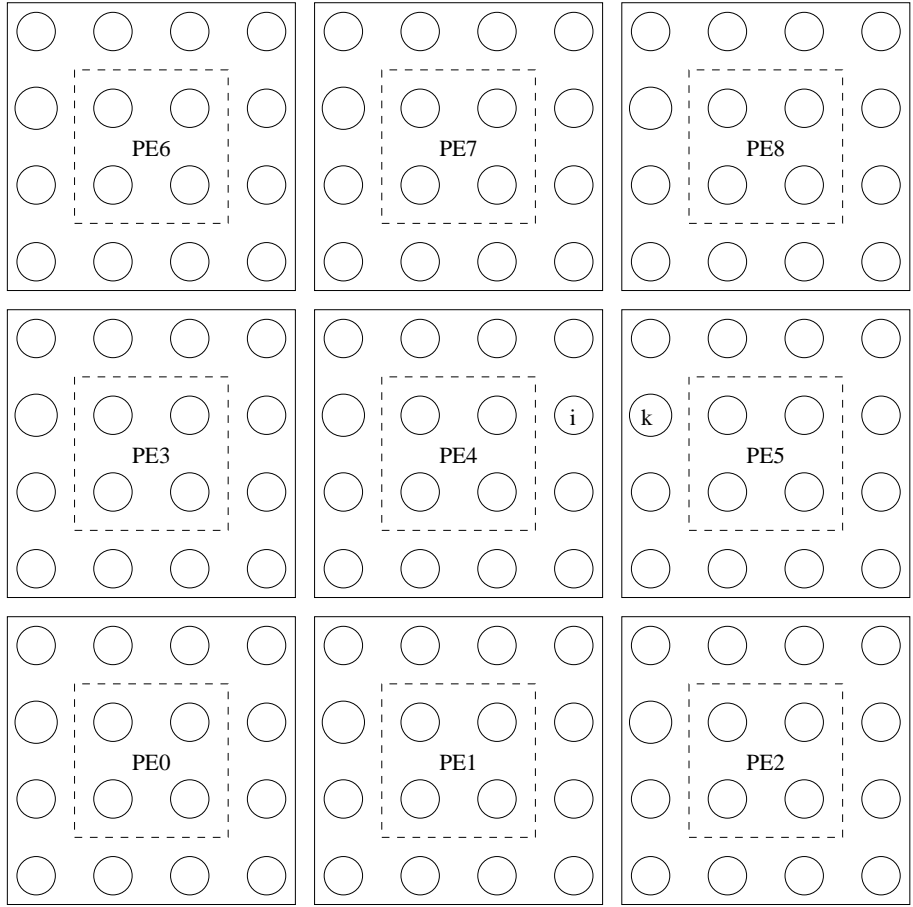}
\end{center}
\caption{Schematic diagram of the spatial decomposition of the system and its
mapping onto a parallel machine. Here $L$$=$$12$ and $l$$=$$4$. Each of the 
$N_{\rm PE}$$=$$(L/l)^{2}$$=$$9$ processing elements (PEs) carries 
$l^2$$=$$16$ spins, confined by solid lines. The spins on the boundary are 
separated from those in the kernel by dashed lines.}
\label{par_map}
\end{figure}

First we review the conservative scheme for parallelization of the 
standard Metropolis algorithm \cite{Luba}. An obvious way to parallelize 
the serial Metropolis algorithm is to spatially 
decompose the $L$$\times$$L$ lattice into $l$$\times$$l$ blocks.
When mapping it onto a parallel computer, each processing element (PE) carries
an $l$$\times$$l$ block of spins, and the number of PEs is $N_{\rm PE}=(L/l)^2$
(Fig.\ \ref{par_map}). Here we  outline the algorithmic steps for the 
continuous-time case. Each PE carries its own {\em local} time $t$. Initially
a spin configuration, corresponding to $t=0$, is chosen, and the time of the
first update is determined by $t = -\ln(r)$, independently on each PE. 
In our case, the initial configuration is $s_i$$=$$+1$ for all spins. 
For clarity, our time unit will be one MCS per PE (MCSP), during which 
each PE attempts to update one spin on average. Each PE is responsible for 
updating the $N=l^2$ spins that it carries by iterating the following steps:
\begin{enumerate}
\item Select a spin from the block with equal probabilities. 
\item
\begin{enumerate}
\item If the chosen spin is in the kernel of the block, go to \mbox{Step 3.}
\item If the chosen spin is on the boundary of the block,
 {\em wait until} the {\em local} simulated time $t$ of the next update
 becomes less than or equal to the same quantity for the corresponding 
 neighboring PE(s) (for our model on a square lattice, $2$ PEs for corner 
 spins,  one PE otherwise), then proceed to \mbox{Step 3.} 
\end{enumerate}
\item Update the state of the spin using the same probabilities as in the 
 standard serial algorithm [Eq.\ (\ref{metropolis})].
\item Determine the time of the new next update by using the {\em local} time 
increment \mbox{$\Delta t = -\ln(r)$}, where $r$ is a uniformly distributed 
random number on $(0,1)$.
\end{enumerate} 

With reliable random number generators (an independent one on each PE), and 
using a continuous probability density (exponential) for the 
time increments, the probability of equal-time nearest-neighbor updates is 
of measure zero. 
Thus, no {\em global} barrier synchronization is necessary among 
the PEs to preserve the uniqueness of the simulated trajectory, provided the 
same set of random seeds are used initially. The algorithm is obviously
free from deadlock, since in the worst-case scenario the PE with the minimum 
local time is able to make progress. It 
is clear from the above {\em asynchronous} algorithm that at any given (wall
clock) moment, different PEs have different local simulated times.
The ``wait until'' directive in Step 2., however, ensures that the 
information passed between neighboring PEs does not violate causality. This
can be seen from the following example. Suppose that PE4 is currently
updating spin $s_i$, which is on the boundary of its block 
(Fig.\ \ref{par_map}). This update is possible if the local update time on PE4,
$t_4$, is not larger than that on PE5, $t_5$. The update is also necessarily 
correct since $s_k$ is guaranteed to be in the same state that it was in at
$t_4$. This is so because updates for PE5 have been forbidden by the ``wait 
until'' directive once $t_5$ became larger than $t_4$.

The above asynchronous algorithm is suitable for a continuous-time
update scheme, but it can cause inconsistency when integer time is used.
Then, explicit barrier synchronization should be incorporated (synchronous 
algorithm) to treat equal-time spin-flip trials of nearest-neighbor spins 
without ambiguity, and to ensure the reproducibility of a simulated path, 
provided the same set of random seeds is used. However, these 
(independent) nearest-neighbor, equal-time events violate detailed balance 
[with respect to the Hamiltonian (\ref{Ising})], 
and are unprecedented in the corresponding serial algorithm.
Thus, we conclude that to be faithful to the original kinetic
Ising model, one must employ the {\em continuous-time} update scheme. Other, 
more general cellular automata may tolerate nearest-neighbor, equal-time 
updates, since in such cases the microscopic update rates are not necessarily 
related to an {\em a priori} known equilibrium probability distribution, and 
the corresponding equal-time events may very well represent the real physical
behavior. Also note that frequent barrier synchronization
can be very ``expensive'' on a distributed-memory architecture, such
as the T3E, where  each PE executes the code completely independently of all 
other PEs. This is especially true for Monte Carlo algorithms, in which the 
core of the update routine is extremely simple and the time a PE spends at a 
barrier waiting for the others can easily exceed the time it is active.

\subsection{The ``shielded'' parallel $n$-fold way algorithm}
A natural idea to parallelize the $n$-fold way algorithm is to employ the
same spatial decomposition as for the parallel Metropolis algorithm, and to 
apply the serial rejection-free update scheme directly on each block.
However, one cannot simply run a copy of the serial $n$-fold way algorithm on 
each PE: due to the nearest-neighbor interaction, the local time increments 
and the class populations
depend not only on the spin values in the block carried by that PE, but also 
on the states of the spins located on the adjacent boundaries of the 
neighboring PEs. By updating a spin on the boundary of a
block, the updating PE could corrupt the simulated history of a 
neighboring block, if the local time of the latter is already ahead of that of 
the former. Thus, the neighboring PE would need to perform a rollback 
procedure to recover from the simulation of a series of ``false'' events. 
It is easy to see that this 
rollback might generate a cascade of rollbacks in other PEs, making the
implementation rather difficult. Moreover, it is unclear how frequently such 
rollbacks would occur, and how far they would go back in time and propagate 
through space. 

Lubachevsky proposed a way to avoid 
rollbacks by modifying the original algorithm \cite{Luba}. In each block,
an additional class is defined which contains the spins on the boundary. The
weight of this class is the number of boundary spins, $N_{\rm b}=4(l-1)$, 
which does not change during the simulation. The original $n$-fold way
tabulation of spin classes is only used in the kernel of the block.
Hence, \mbox{$N_{\rm b}+\sum_{i=1}^{10} n_i =N$}, where 
$\sum_{i=1}^{10} n_i=N_{\rm k}$ is the
number of spins in the kernel, and $N=l^2$ is the total number of 
spins in a block. Note that the population of the classes,  
$\{N_{\rm b},\{n_i\}_{i=1}^{10} \}$, is maintained separately for each block by
the corresponding PE. Once the initial configuration is set and the 
corresponding local time of first update is determined, the asynchronous 
algorithm (with continuous time) consists of the following iterated steps 
performed by each PE:
\begin{enumerate}
\item Select a class according to the relative weights 
$\{N_{\rm b},\{n_i p_i\}_{i=1}^{10} \}$, and select a spin from the chosen 
class with equal probabilities.
\item
\begin{enumerate}
\item If the chosen spin is in the kernel, flip it with probability one and
go to \mbox{Step 3.} 
\item If the chosen spin belongs to the boundary class, 
{\em wait until} the {\em local} simulated time of the next update 
becomes less than or equal to the same quantity for the corresponding 
neighboring PE(s). Then the state of this spin {\em may or may not} change: 
flip it with the usual Metropolis probability [Eq.\ (\ref{metropolis})] and
proceed to \mbox{Step 3.} 
\end{enumerate}
\item Update the tabulation of the spin classes $\{n_i\}_{i=1}^{10}$ in the 
kernel.
\item Determine the time of the new next update (in units of MCSP) by using 
the {\em local} time increment, \mbox{$\Delta t=-\ln(r)/\Gamma_s\;,$} where
\begin{equation}
\Gamma_s=\frac{N_{\rm b} + \sum_{i=1}^{10} n_i p_i}{N}\;,
\label{increment}
\end{equation}
and $r$ is a uniformly distributed random number on $(0,1)$.
\end{enumerate}
Thanks to the introduction of the class of boundary spins (which has 
{\em fixed} weight), 
the neighboring PEs are shielded from each other, since a spin flip 
on the boundary of a block does not effect the tabulation of spin classes in 
an adjacent one. Hence, just like in the standard parallel scheme, the same 
conservative approach (the ``wait until'' directive in \mbox{Step 2.}) ensures 
that the information passed between PEs is valid and the updates are correct.

Again, (employing proper synchronization) one can experiment with integer-time
updates. Although in this case time increments are random integers and the 
probability of picking two nearest-neighbor spins residing on two adjacent PEs
with equal update times is small, it is nevertheless nonzero. We have 
already argued that these (perhaps rare) events have no corresponding 
analogues in the serial algorithm. 
Furthermore, the use of barrier synchronization severely 
degrades performance. Although, for ``experimentation'' purposes on the 
architecture we implemented the integer-time, synchronous algorithm as well 
\cite{int_nfold}, we will not discuss its performance in detail here.

\section{Performance}
We implemented both the parallel Metropolis and the parallel $n$-fold way 
algorithms (as described in the previous section) on the Cray T3E parallel 
architecture at the National Energy Research Scientific Computing Center 
(NERSC). For message passing, we employ the Cray-specific, 
logically shared, distributed memory access (SHMEM) routines. 
The fast SHMEM library supports communication initiated by {\em one} 
PE, together with remote atomic memory operations. Without these features, it 
would be extremely inconvenient to code an algorithm for stochastic simulation
on a distributed memory machine, where the pattern of communication is 
completely unpredictable. These characteristics clearly outweigh the loss of 
portability of our code.

To better understand the performance of our implementation we monitored the
following quantities:

\paragraph*{Utilization.} This is defined as the fraction of ``non-idling'' 
PEs, i.e., the ones which either pick a spin in the kernel or successfully 
pass the ``wait until'' directive in Step 2. of the algorithms. Since the 
routines are asynchronous 
(the main simulation cycles on each PE are not executed in lock-step) it is 
fairly difficult to measure this ratio. To obtain an 
estimate, we placed explicit barrier synchronization in our code and 
performed {\em separate} runs. The performance was irrelevant for these runs, 
the only information that we aimed for was the fraction of non-idling PEs in 
each (now artificially lock-step) main simulation cycle. We emphasize that the
utilization only measures the fraction of PEs that are not idle. It does
{\em not} say anything about whether the active PEs are doing anything
``useful'' in the sense of performing successful updates.

\paragraph*{The mean local time increment, $\overline{\Delta t}$.} Although 
$\Delta t$ in the $n$-fold way algorithm is not stationary over the course of 
the metastable decay, its mean carries important information about how 
successfully the algorithm ``bypasses'' those spin-flip attempts which would
be rejected if the Metropolis algorithm were used. For the parallel Metropolis
algorithm $\overline{\Delta t}=1$ trivially ($\overline{\ln(r)}=1$). 

\paragraph*{PE update rate.} This quantity is literally the simulation speed 
of a PE in units of standard MCS per PE per second, i.e., MCSP/s. This is 
an ``absolute'' measure and in fact tells which parallel algorithm should 
be used for optimal simulation speed. The full update rate of the parallel 
algorithm is simply $\mbox{(PE update rate)}$$\times$$N_{\rm PE}$.

In order to compare directly the performances of the parallel algorithms 
to those of their serial counterparts, we also ran the corresponding serial 
routines on one node of the T3E and determined the following measures:

\paragraph*{Efficiency.} This is the ratio of the PE update rate of
the parallel algorithm to the update rate of the corresponding serial 
algorithm using the same full system size $L$. For both parallel algorithms
it is proportional to the utilization, and it is related to the communication 
speed of the architecture through the fraction of spin-flip attempts in which
message passing is needed. For our algorithms, due to the fast communication 
hardware of the architecture, the communication overhead is a small effect 
compared to the inherent low utilization, which is a common drawback of 
conservative parallel methods. For the parallel $n$-fold way algorithm the 
efficiency is also proportional to the ratio of the typical time increments of
the parallel and the corresponding serial routine, 
$\overline{\Delta t}_{\rm par}/\overline{\Delta t}_{\rm ser}$.

\paragraph*{Speedup.} This is the ratio of the (full) update rate of 
the parallel algorithm to the update rate of its corresponding serial 
counterpart, i.e., $\mbox{efficiency}$$\times$$N_{\rm PE}$.

Both efficiency and speedup are {\em relative} measures, and they merely 
indicate how successfully the parallelization of the {\em corresponding} 
serial algorithm is accomplished. As we shall see, a parallel code with 
lower efficiency can outperform one with higher efficiency, due to the faster 
``serial'' core of the former. For very large systems, 
direct simulation using the serial algorithms is not possible, due to memory 
limits. The largest system sizes we could allocate were $L$$=$$2048$ for the 
serial Metropolis and $L$$=$$1280$ for the serial $n$-fold way algorithm. To 
obtain speedup and efficiency estimates for larger systems we extrapolated 
the smaller-system update rates of the serial routines. 

Before discussing the performance of the ``shielded'' $n$-fold way algorithm, 
we note two inherently weak features, which are not related to the otherwise 
fast communication hardware of the T3E parallel architecture.
  
First, as a general guideline, the fewer communications one has to execute, 
the better is the performance of the parallel code. In our case, the 
probability of picking a spin on the boundary, which will be followed 
by some kind of communication between neighboring PEs, is greater than the 
surface-to-volume ratio, $N_{\rm b}/N=4(l-1)/l^{2}\approx 4/l$. 
In particular, it is determined by the relative 
weights in the modified $n$-fold way algorithm, which are given by 
\mbox{$N_{\rm b}/(N_{\rm b}+\sum_{i=1}^{10} n_i p_i)$}. 
With very small $p_i$'s this ratio can take unfavorably large values, close 
to unity, leading to more frequent message passing and, more importantly, 
idling if required by the ``wait until'' directive. 

Second, the typical time increment of the parallel algorithm is smaller than 
for the serial one. As mentioned  in Section II, the advantage of the serial 
$n$-fold way routine lies in the large typical time increments that correspond
to the very small flipping probabilities at low temperatures. However, for
the same sets of class-specific flipping probabilities $p_{i}$, the mean time 
increments of the  ``shielded'' parallel and serial $n$-fold way routines are
related as
\begin{equation}
\frac{1}{\overline{\Delta t}_{\rm par}}\approx
\frac{N_{\rm b}}{N} + \frac{\overline{\sum_{i=1}^{10} n_i p_i} }{N_{\rm k}} 
\frac{N_{\rm k}}{N} \approx
\frac{N_{\rm b}}{N} + \frac{1}{\overline{\Delta t}_{\rm ser}} 
\frac{N_{\rm k}}{N} =
\frac{1}{\overline{\Delta t}_{\rm ser}} + 
\frac{N_{\rm b}}{N}(1-\frac{1}{\overline{\Delta t}_{\rm ser}})\;.
\label{inc_degrade}
\end{equation}
This follows from Eq.\ (\ref{increment}) and implies that 
$\overline{\Delta t}_{\rm par} < \overline{\Delta t}_{\rm ser}$ always.
Thus at very low temperatures, where 
\mbox{$\overline{\Delta t}_{\rm ser} \gg l/4$, $\overline{\Delta t}_{\rm par}$}
is determined almost completely by the block size, rather than by the $p_{i}$'s
and the populations of the corresponding classes:
\begin{equation}
\overline{\Delta t}_{\rm par}\stackrel{<}{\sim} l/4 \;.
\label{inc_par}
\end{equation}

We test the scaling of the parallel algorithms up to 400 PEs. 
First, the system size is kept constant, and we divide it into 
smaller and smaller blocks (Fig.\ \ref{fixed_L}). Then, alternatively, 
we keep the block size fixed and study the scaling for larger and larger
systems by increasing the number of blocks (Fig.\ \ref{fixed_l}). We also 
carry out experiments with fixed number of PEs and varying 
block size (Fig.\ \ref{fixed_pe}) to study in detail the effect of 
increasing surface-to-volume ratio of the blocks. Each of these studies were
performed at $T$$=$$0.7T_{c}$ and $|H|/J$$=$$0.2857$, where the typical 
droplet separation, $R_{o}$, is approximately $32$ lattice constants. 
Finally, with fixed number 
of PEs and fixed block size we study the effect of the temperature and 
magnetic field on the performance (Fig.\ \ref{var_TH}), to determine the 
regime of efficient applications to our particular model system. 
The results reflect the features discussed in the previous paragraph.

\subsection{Scaling with $N_{\rm PE}$ for fixed system size.}
\begin{figure}[t]
\begin{center}
\epsfxsize=7.0cm\epsfysize=7.0cm\epsfbox{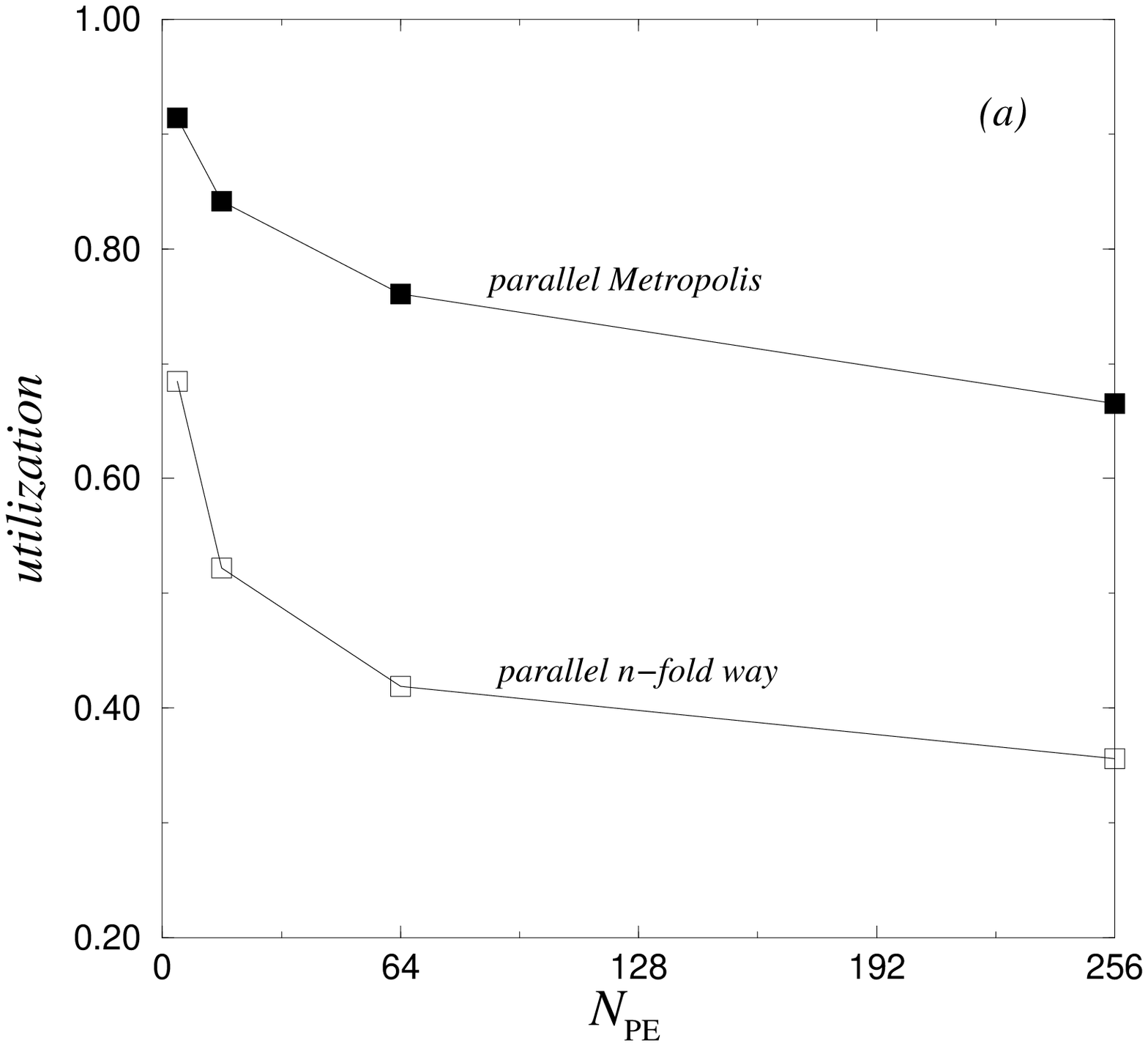} \hspace*{1cm}
\epsfxsize=7.0cm\epsfysize=7.0cm\epsfbox{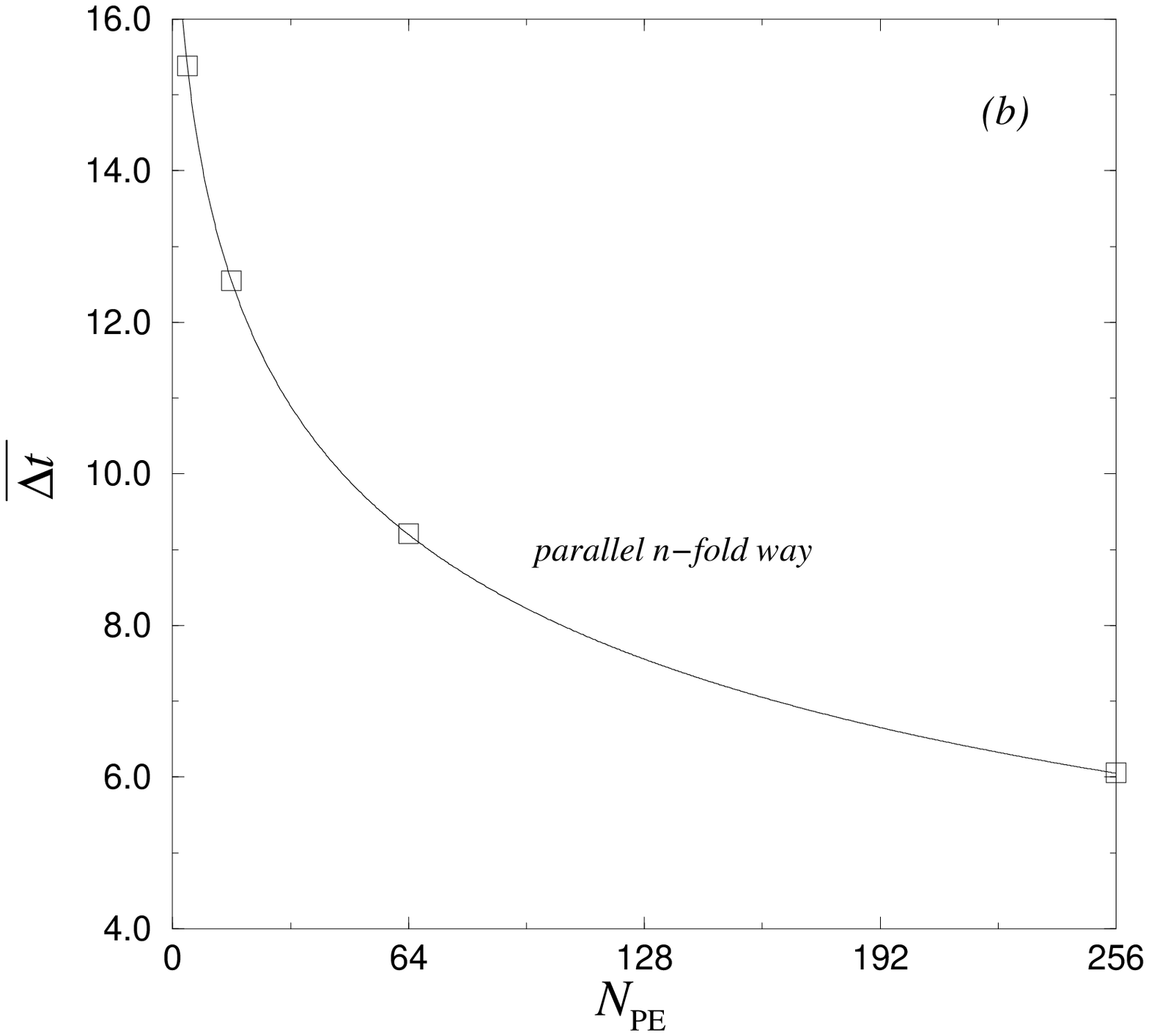} \\ \vspace*{1cm}
\epsfxsize=7.0cm\epsfysize=7.0cm\epsfbox{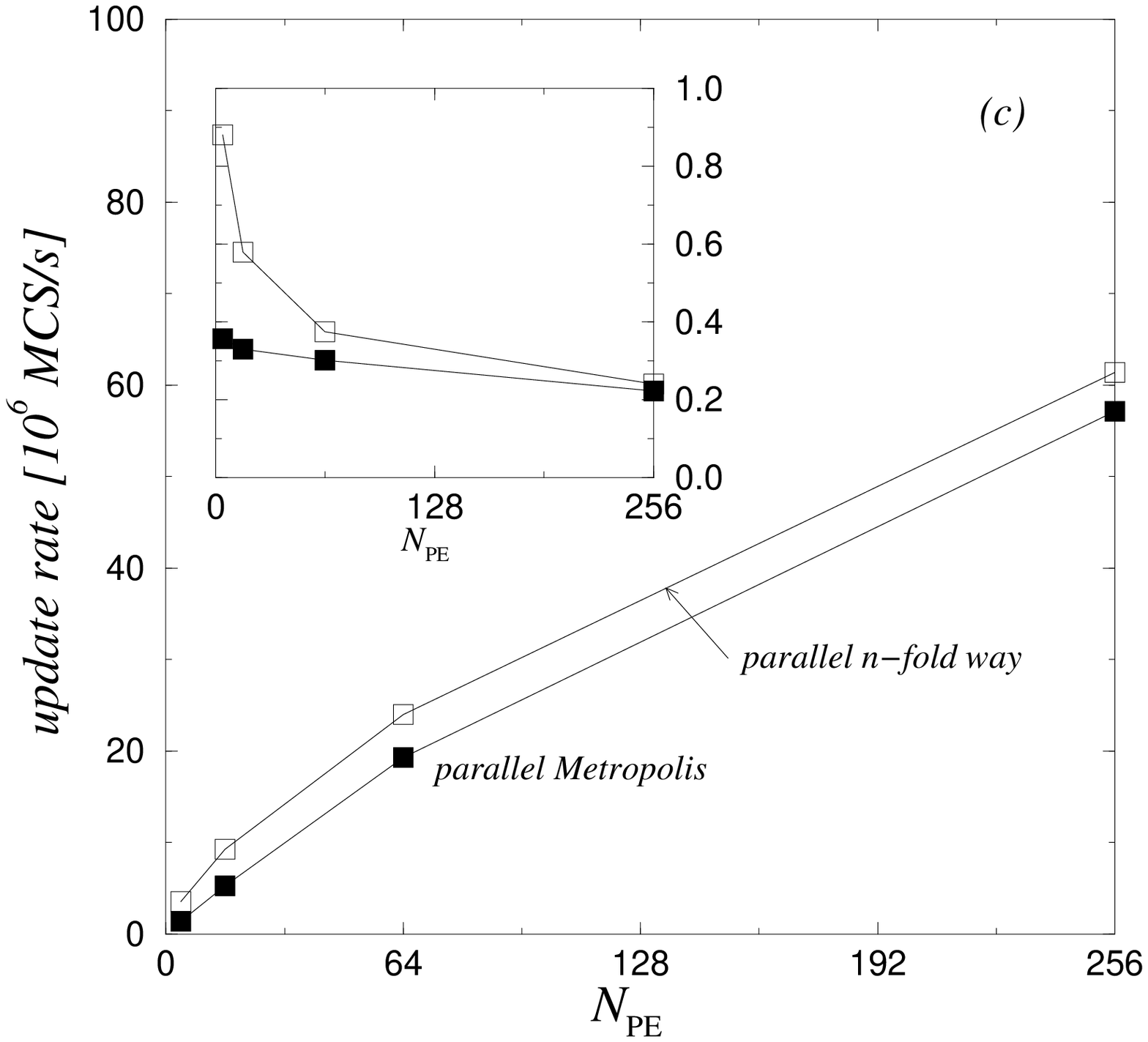} \hspace*{1cm}
\epsfxsize=7.0cm\epsfysize=7.0cm\epsfbox{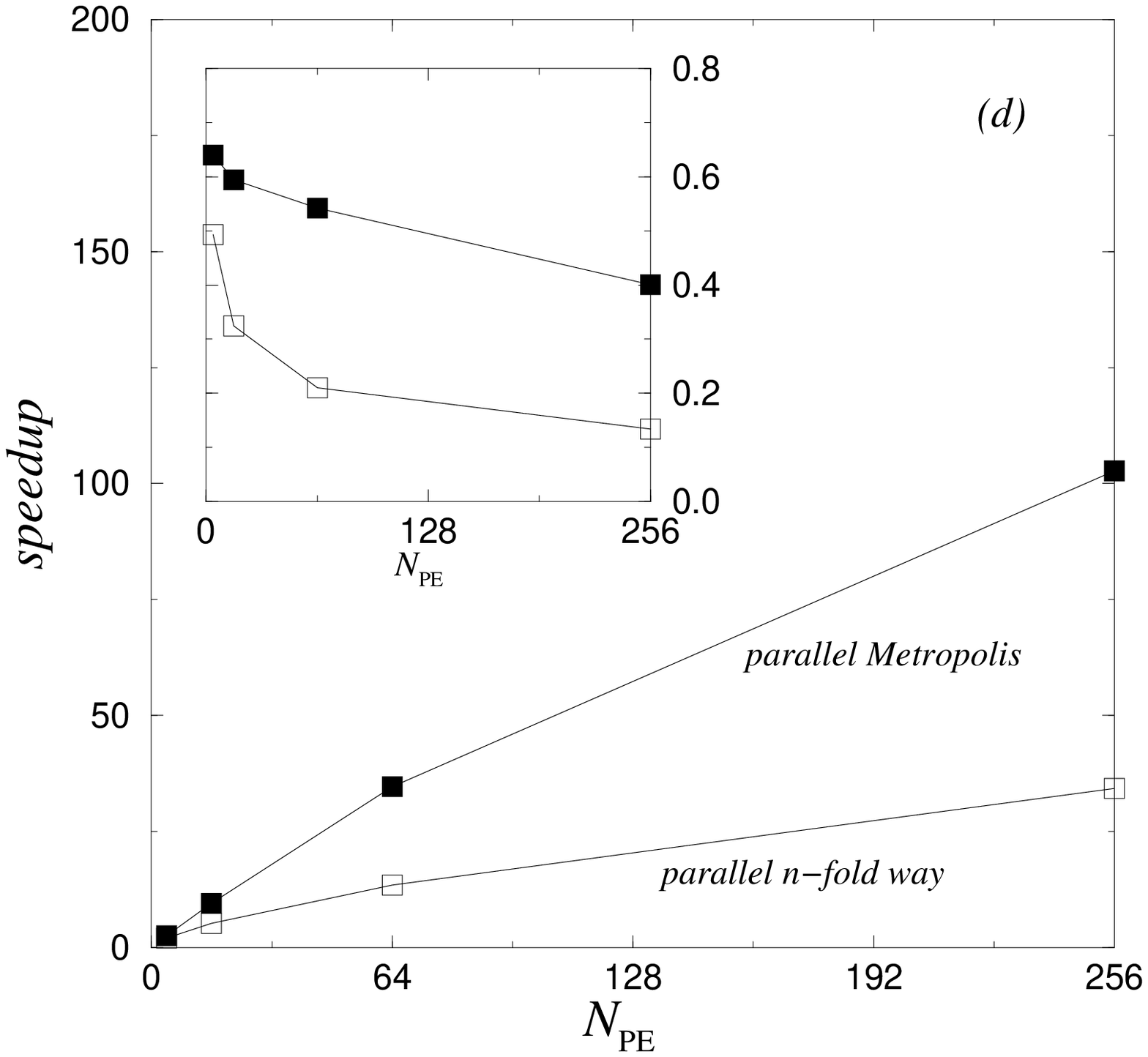}
\end{center}
\caption{Scaling and performance analysis for fixed system size, $L$$=$$512$, 
at $T$$=$$0.7T_{c}$ and $|H|/J$$=$$0.2857$, comparing the parallel $n$-fold 
way algorithm (open squares) and the parallel Metropolis algorithm (filled 
squares). The lines connecting the data points are merely guides to the eye 
except in (b), where they represent the theoretical prediction of Eq.\ 
(\protect\ref{inc_degrade}).
(a) Utilization. 
(b) Mean time increment, $\overline{\Delta t}$, for the parallel $n$-fold way
algorithm. 
(c) Update rate (inset: PE update rate in units of $10^6$ MCSP/s). 
(d) Speedup (inset: efficiency).}
\label{fixed_L}
\end{figure}

For this series of runs we choose $L$$=$$512$ for the total system size and 
we employ $N_{\rm PE}$$=$$4, 16, 64$, and $256$ (corresponding to 
$l$$=$$256, 128, 64$, and $32$, respectively). 
For both the parallel $n$-fold way and the parallel Metropolis algorithm the
utilization drops with decreasing block size (Fig.\ \ref{fixed_L}a). The 
utilization for the $n$-fold way routine is significantly lower than for
Metropolis: the probability of choosing a spin on the boundary is greater than 
the surface-to-volume ratio and then it is ultimately followed by an inquiry 
of the local time of the neighbor(s) and possible idling. Further, the typical
time increments are decreasing as well (Fig.\ \ref{fixed_L}b). Note that for 
the {\em serial} $n$-fold way routine the mean time increment is 
$\overline{\Delta t}_{\rm ser}$$=$$19.9$ (Table \ref{table1}). This is the only
parameter in Eq. (\ref{inc_degrade}), which gives complete agreement with the
measured values of $\overline{\Delta t}_{\rm par}$, as shown by the solid
curve in Fig.\ \ref{fixed_L}b. As a result of 
these factors, the PE update rate of the parallel $n$-fold algorithm drops 
faster than that of the parallel Metropolis one (Fig.\ \ref{fixed_L}c inset). 
For sufficiently small blocks, the performance of the $n$-fold routine 
actually becomes poorer than for the Metropolis routine, as we shall see in 
the experiments with fixed number 
of PEs and varying block size. For both algorithms the scaling is 
systematically {\em worse} than linear, for the reasons explained above 
(Fig.\ \ref{fixed_L}c). In particular, the parallel $n$-fold way for 
large number of PEs (small block size $l$) cannot scale better than 
$\sqrt{N_{\rm PE}}$,  since for fixed $L$ the typical time increment scales as 
$\overline{\Delta t}_{\rm par}\stackrel{<}{\sim} l/4 \sim 1/\sqrt{N_{\rm PE}}$. 
Although employing many {\em small} blocks does result in speedup, the 
efficiency clearly becomes poorer at the same time (Fig.\ \ref{fixed_L}d).  

For completeness, we mention that the performance
of the integer-time, synchronous $n$-fold way routine becomes progressively
weaker than that of the asynchronous version with continuous time 
for increasing number of PEs.  For example, with $256$ PEs it runs $2.3$ 
times slower than the continuous-time routine, due to the necessity of 
explicit barrier synchronizations. It is even slower than the asynchronous 
Metropolis routine by a factor of $2.1$.
 
\subsection{Scaling with $N_{\rm PE}$ for fixed block size.} 
\begin{figure}[t]
\begin{center}
\epsfxsize=7.0cm\epsfysize=7.0cm\epsfbox{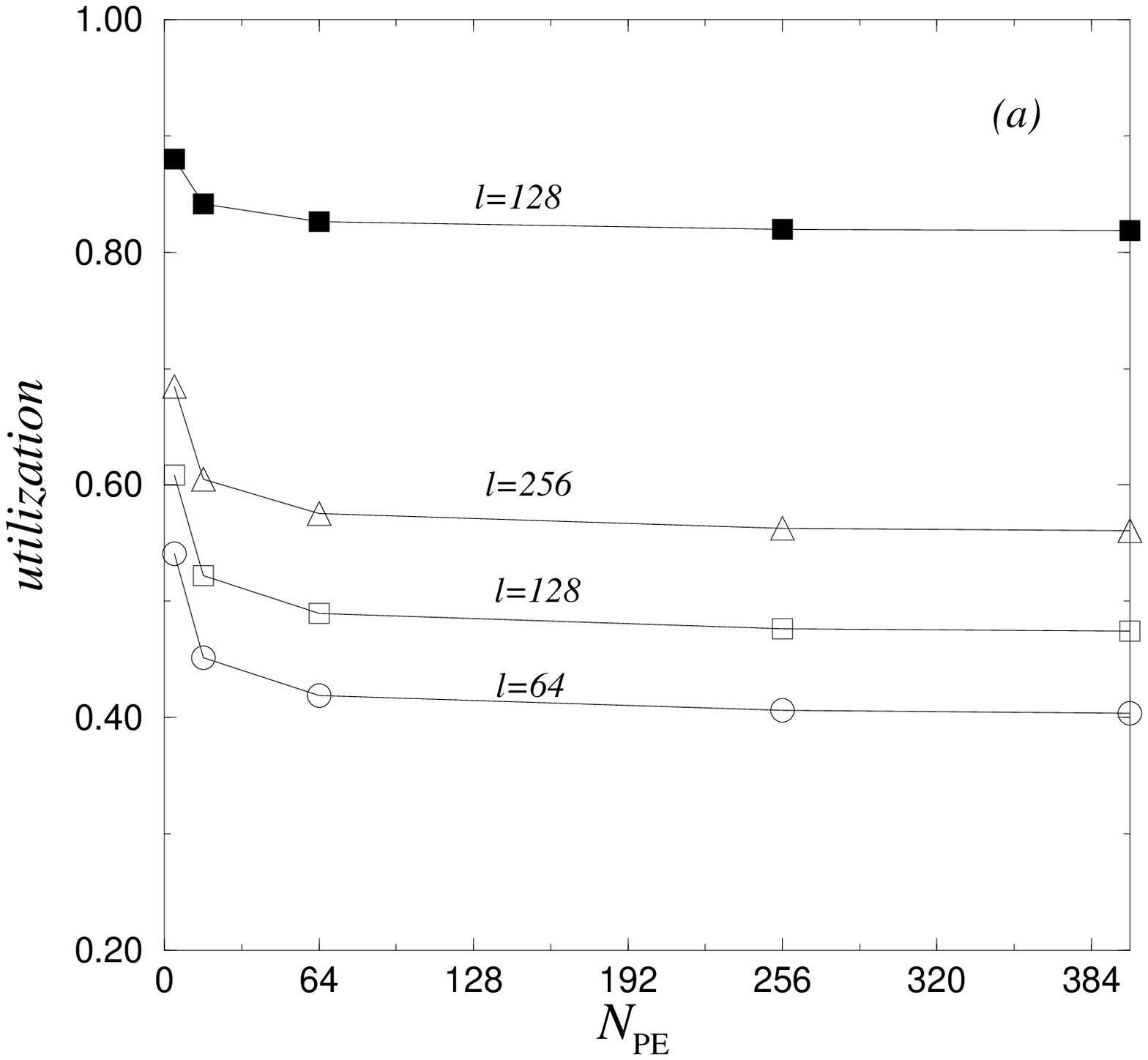} \hspace*{1cm}
\epsfxsize=7.0cm\epsfysize=7.0cm\epsfbox{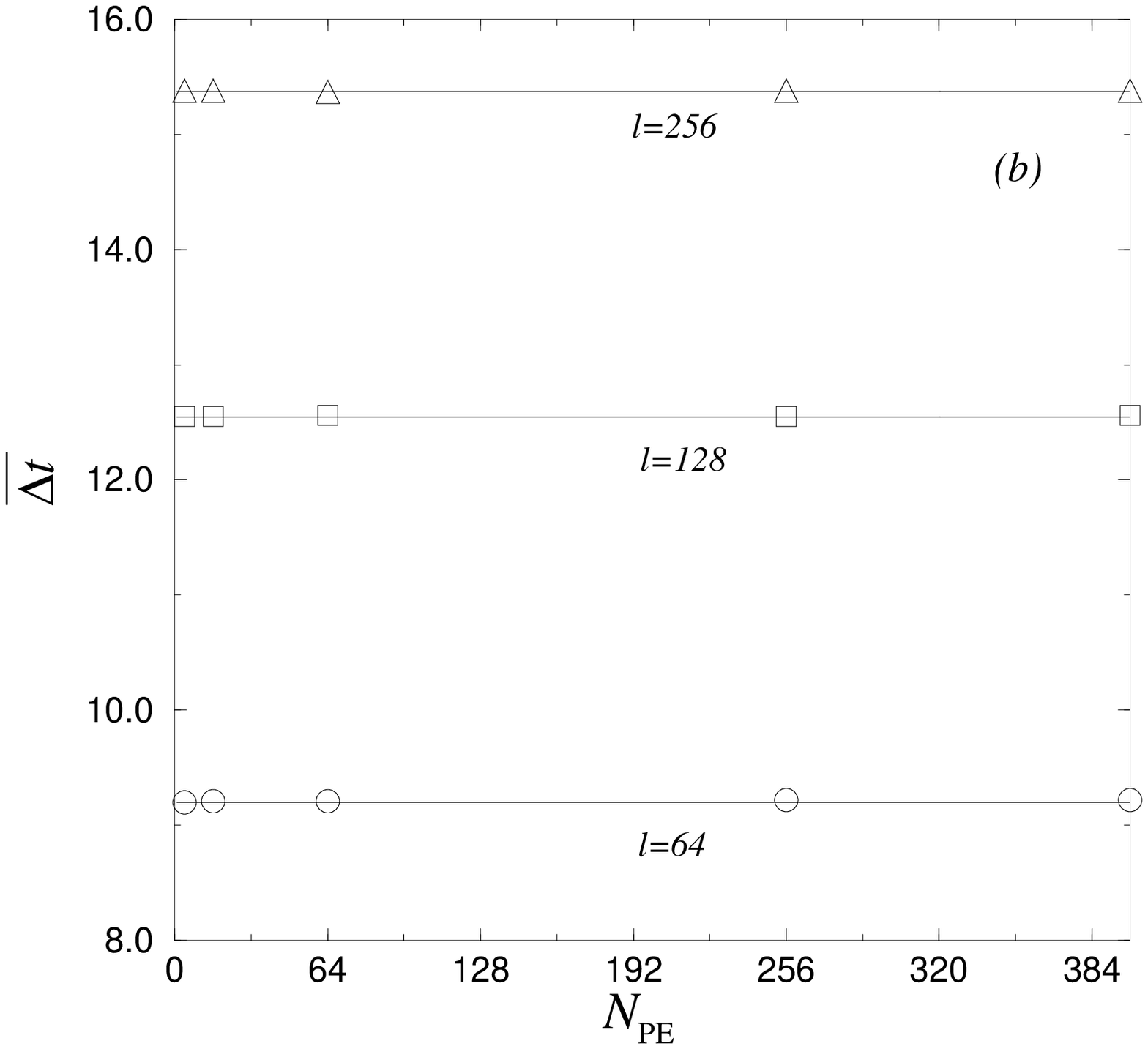} \\ \vspace*{1cm}
\epsfxsize=7.0cm\epsfysize=7.0cm\epsfbox{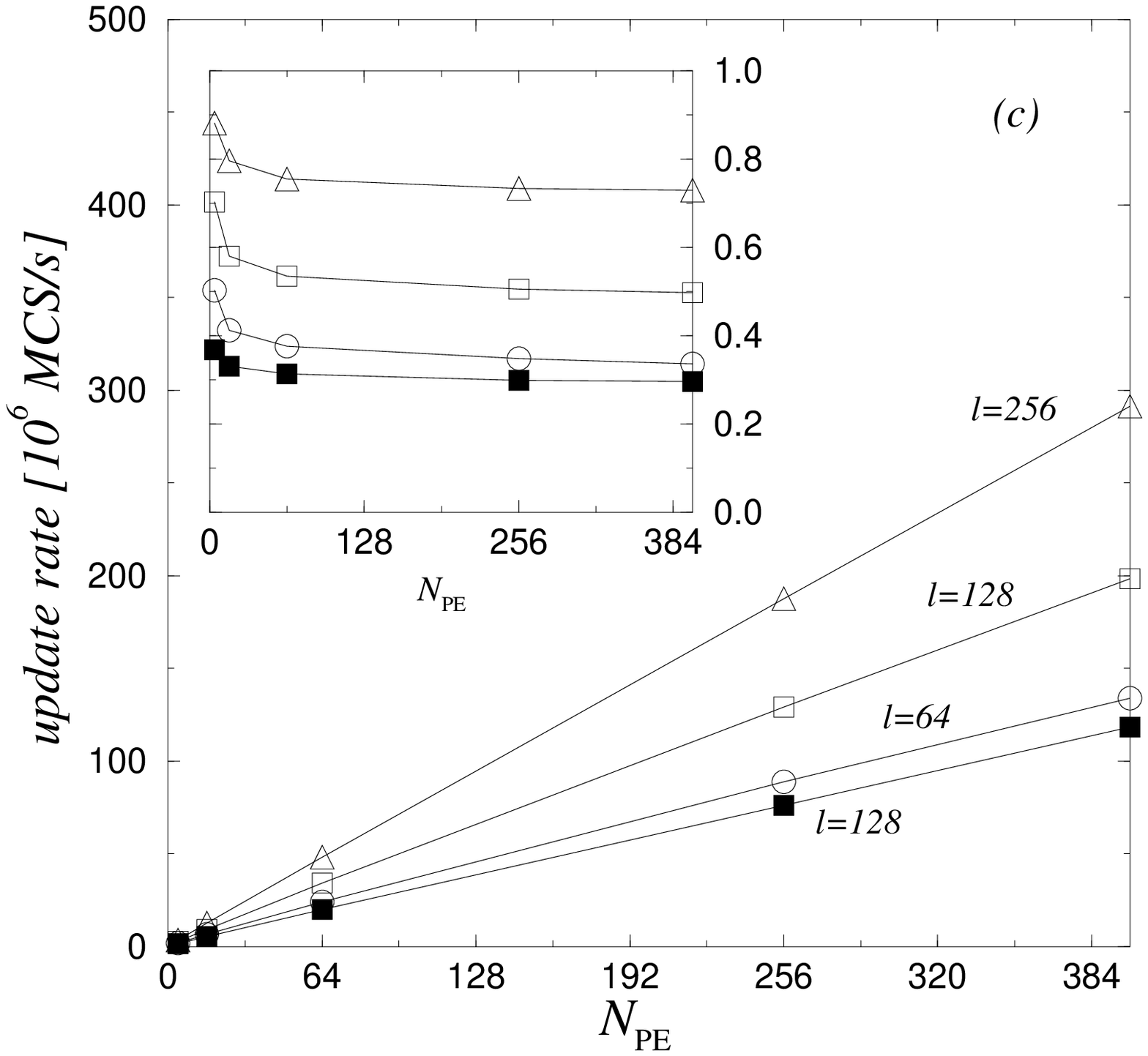} \hspace*{1cm}
\epsfxsize=7.0cm\epsfysize=7.0cm\epsfbox{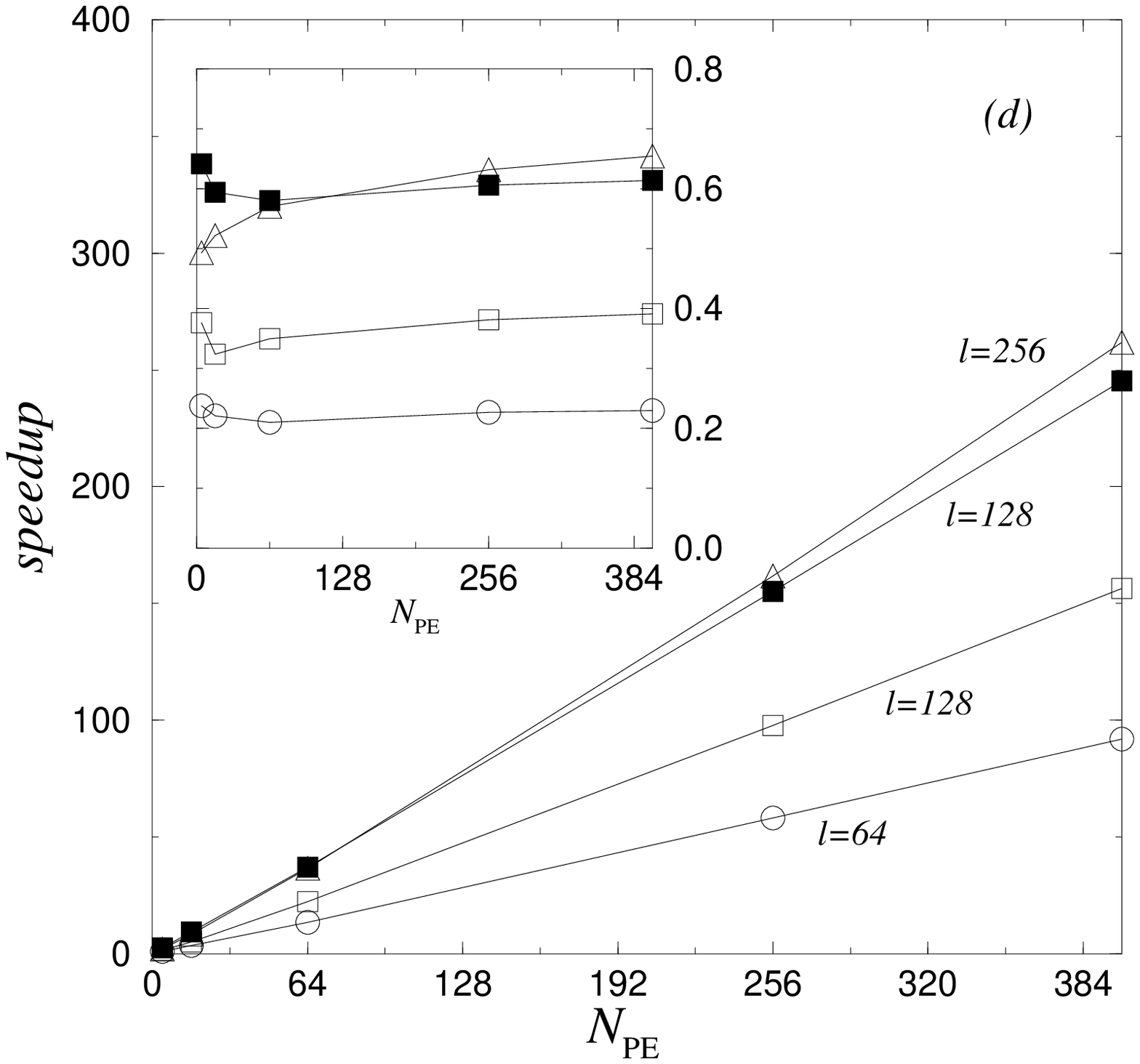}
\end{center}
\caption{Scaling and performance analysis for three different block sizes, 
$l$$=$$64, 128$, and $256$ for the parallel $n$-fold way algorithm (open 
circles, squares, and triangles, respectively), and $l$$=$$128$ for the 
parallel Metropolis algorithm (filled squares) at $T$$=$$0.7T_{c}$ and 
$|H|/J$$=$$0.2857$. The linear system size is $L$$=$$l\sqrt{N_{\rm PE}}$.
The lines connecting the data points are merely guides to 
the eye except in (b), where they represent the theoretical prediction of Eq.\ 
(\protect\ref{inc_degrade}).
(a) Utilization. 
(b) Mean time increment, $\overline{\Delta t}$, for the parallel $n$-fold way 
algorithm. 
(c) Update rate (inset: PE update rate in units of $10^6$ MCSP/s). 
(d) Speedup (inset: efficiency).}
\label{fixed_l}
\end{figure}

Here we keep the 
size of the blocks fixed (using three different values: $l$$=$$64, 128, 256$), 
while increasing the system size by using larger numbers of PEs 
($N_{\rm PE}$$=$$4, 16, 64, 256,\mbox{and}\; 400$). Clearly, the
larger the block size, the higher the utilization. Even with fixed $l$, the
utilization monotonically decreases with increasing $N_{\rm PE}$. However, it 
appears to approach a {\em nonzero} value for large numbers of PEs 
(Fig.\ \ref{fixed_l}a).  As was pointed out in Ref. 
\cite{Luba}, it is a highly non-trivial mathematical problem to prove that
the utilization tends to a nonzero value as $N_{\rm PE}\rightarrow\infty$. 
For practical purposes, given our architecture with 512 nodes this 
question might seem academic, but it truly lies at the heart
of this conservative approach to discrete-event simulation. Again, the 
utilization for the parallel Metropolis algorithm significantly exceeds that 
for the $n$-fold way: for the Metropolis routine with $l$$=$$128$ it is 
$82\%$ with $400$ PEs, while for the $n$-fold way it is only $56\%$, even for
$l$$=$$256$. With fixed $l$, the mean time increments are independent of 
$N_{\rm PE}$ in the parallel $n$-fold way routine (Fig.\ \ref{fixed_l}b). 
However, increasing $l$ systematically improves $\overline{\Delta t}$ and thus 
the performance. We find almost {\em linear} scaling of the update rate with 
$N_{\rm PE}$ for both parallel algorithms (Fig.\ \ref{fixed_l}c,d). Despite 
its relatively low utilization, the $n$-fold way routine clearly outperforms 
the Metropolis one, due to its large time increments (partly rejection-free 
nature).

Again we note that our implementation of the integer-time synchronous 
$n$-fold way algorithm performs rather poorly compared to its asynchronous
counterpart: it runs $2.3$ times slower than
the asynchronous $n$-fold way and $1.3$ times slower than the asynchronous
Metropolis routine with $l$$=$$128$ and $256$ PEs.

\subsection{Simulation with fixed number of PEs and varying block size.} 
\begin{figure}[t]
\begin{center}
\epsfxsize=7.0cm\epsfysize=7.0cm\epsfbox{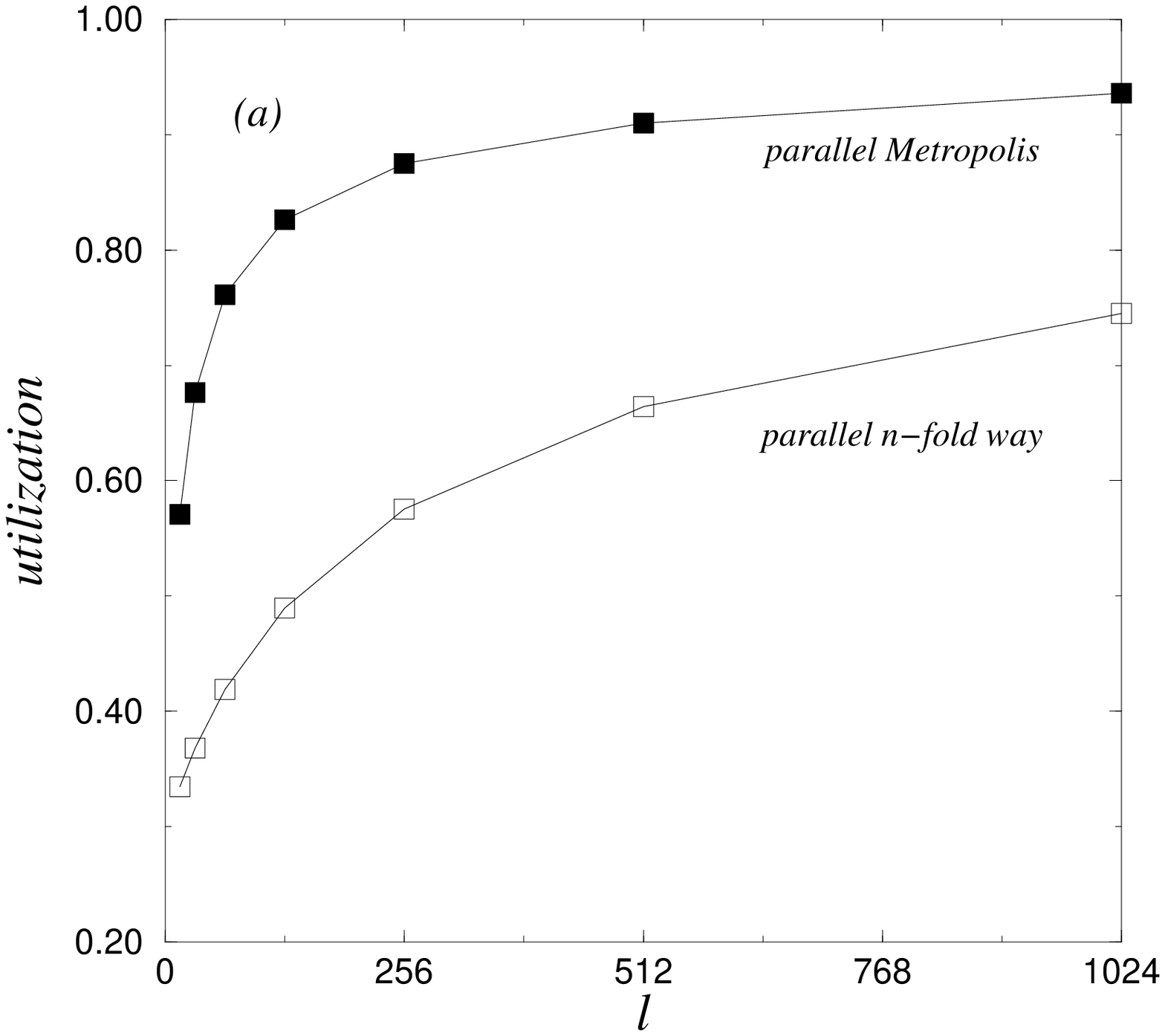} \hspace*{1cm}
\epsfxsize=7.0cm\epsfysize=7.0cm\epsfbox{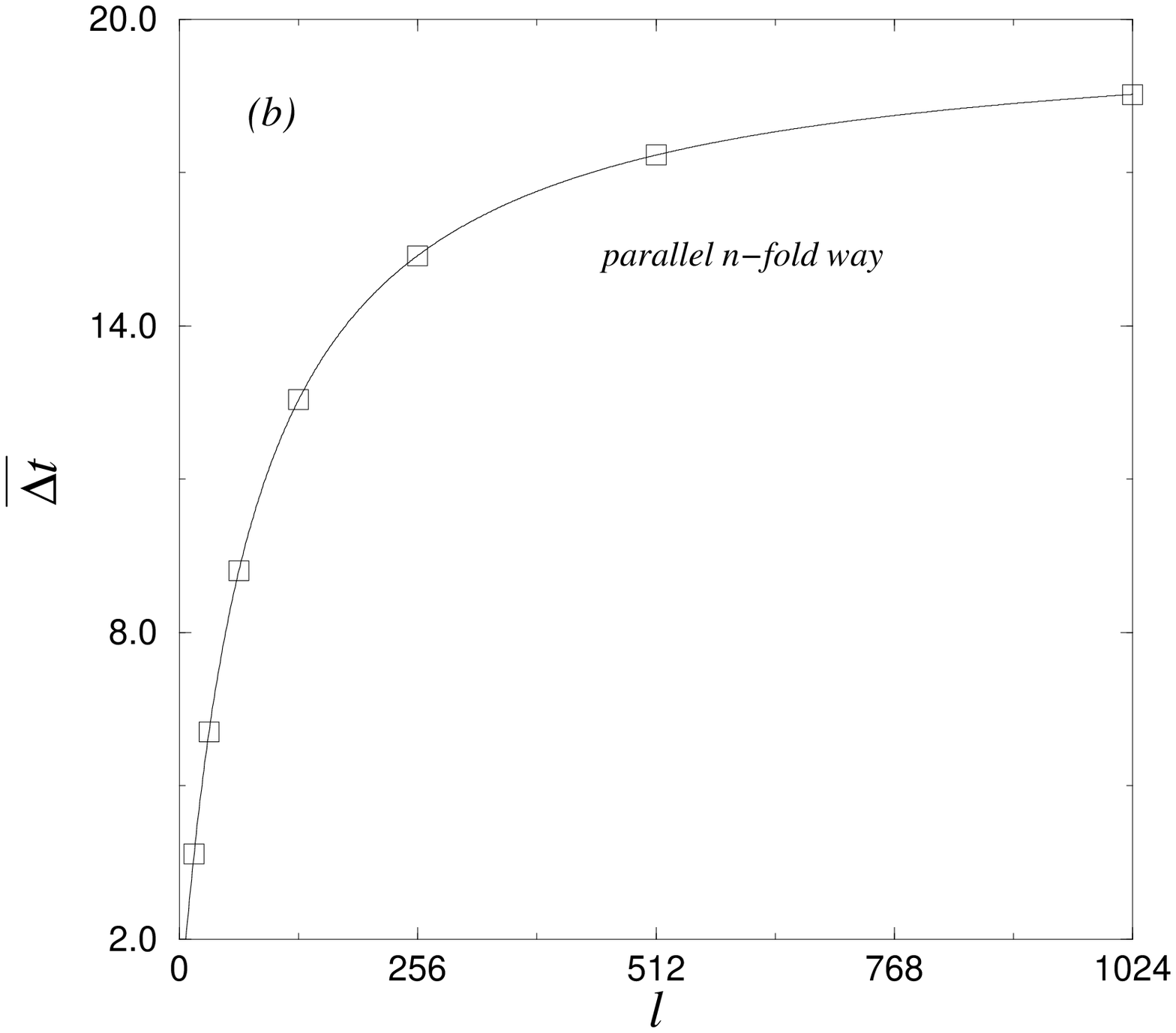} \\ \vspace*{1cm}
\epsfxsize=7.0cm\epsfysize=7.0cm\epsfbox{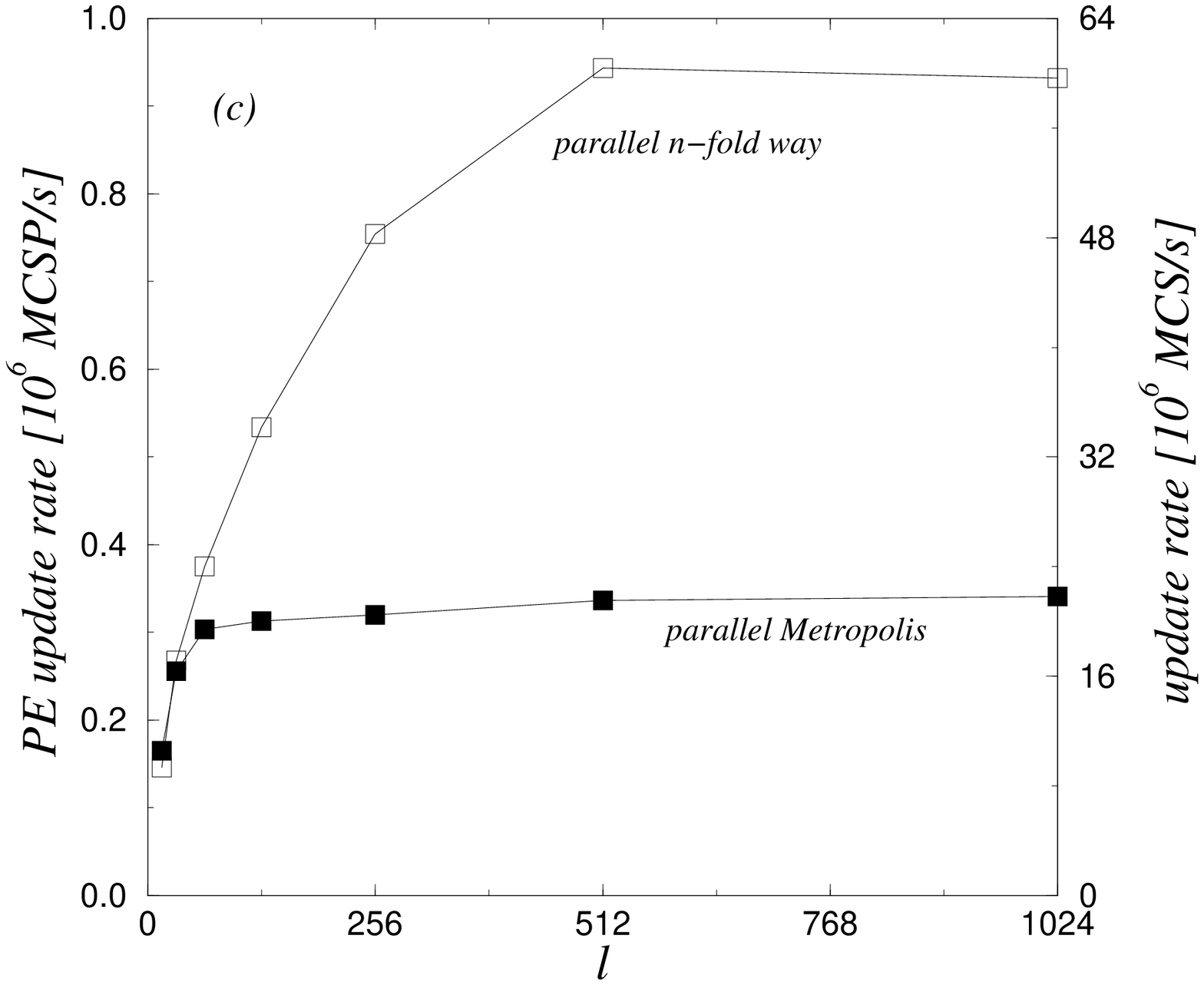} \hspace*{1cm}
\epsfxsize=7.0cm\epsfysize=7.0cm\epsfbox{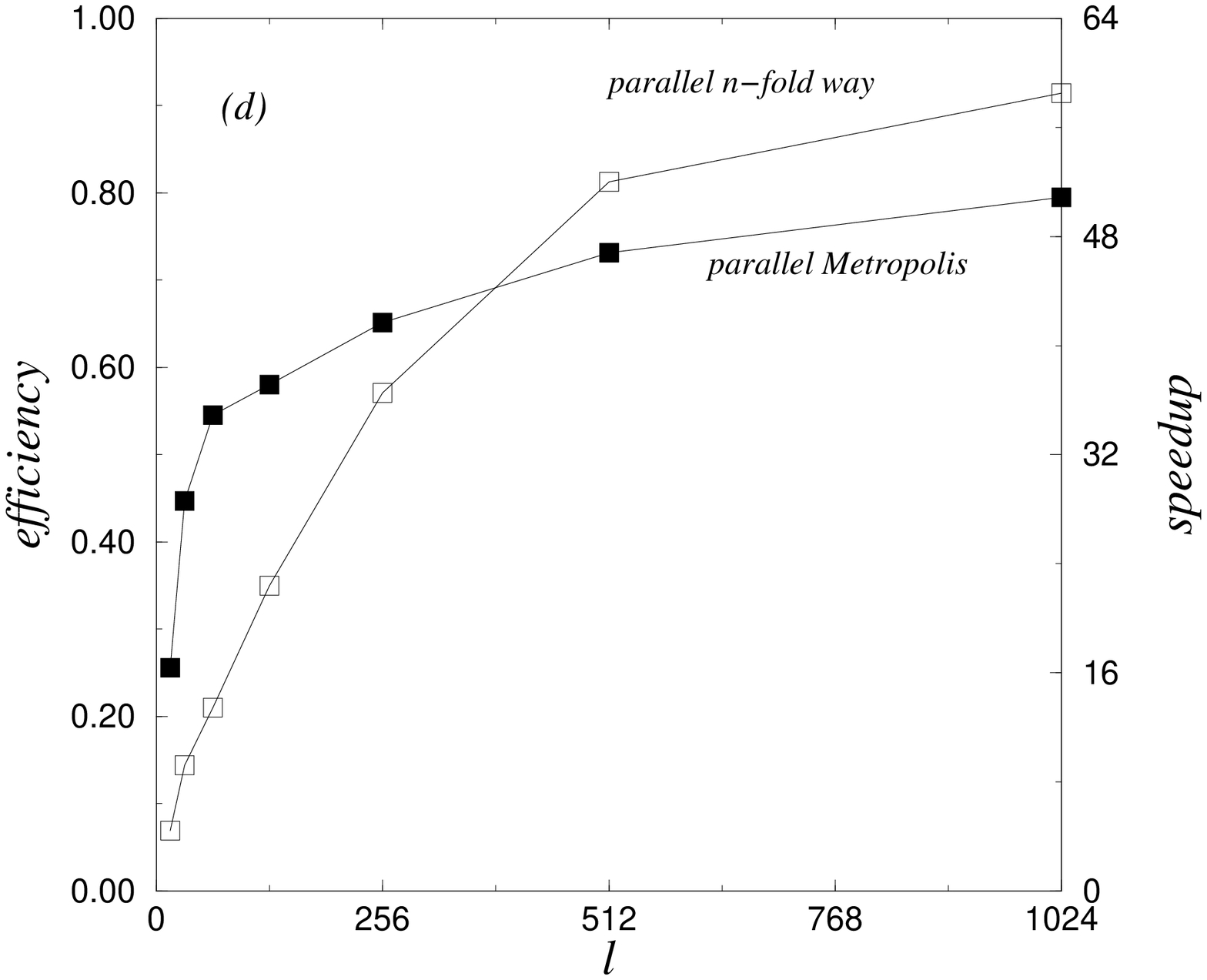}
\end{center}
\caption{Performance analysis for fixed number of PEs, $N_{\rm PE}$$=$$64$, as a 
function of the block size, $l$, 
for the parallel $n$-fold way algorithm (open squares), and for the parallel 
Metropolis algorithm (filled squares) at $T$$=$$0.7T_{c}$ and 
$|H|/J$$=$$0.2857$. The lines connecting the data points are merely guides to 
the eye except in (b), where they represent the theoretical prediction of Eq.\ 
(\protect\ref{inc_degrade}). 
(a) Utilization. 
(b) Mean time increment, $\overline{\Delta t}$, for the parallel $n$-fold way 
algorithm. 
(c) Update rate (right scale) and PE update rate (left scale). 
(d) Speedup (right scale) and efficiency (left scale).}
\label{fixed_pe}
\end{figure}

Here $N_{\rm PE}$$=$$64$, which can be regarded as a reasonable number for 
production runs. We increase the block size ($l$$=$$16, 32, 64, 128, 256, 512,
\mbox{and}\;1024$) and ``stretch'' the routine close to its memory limits.
For both algorithms the utilization increases with increasing block size. This 
happens faster for the parallel Metropolis routine, for which the utilization 
is more directly related to the surface-to-volume ratio of the blocks
(Fig.\ \ref{fixed_pe}a): for $l$$=$$1024$ it is $93\%$ while it is $74\%$ for 
the $n$-fold way routine. Further, the mean time increment for the parallel 
$n$-fold way algorithm systematically approaches that of its serial 
counterpart, $\overline{\Delta t}_{\rm ser}$$=$$19.9$
(Fig.\ \ref{fixed_pe}b and Table \ref{table1}). 
For large $l$  such that $l/4 \gg\overline{\Delta t}_{\rm ser}$, 
$\overline{\Delta t}_{\rm par}\stackrel{<}{\sim} 
\overline{\Delta t}_{\rm ser}$, as expected from Eq. (\ref{inc_degrade}).
The data are in complete agreement with our theoretical result, given by the
solid curve.
Here the temperature and field are moderate, so that the increasing
block size $l$ allows $\overline{\Delta t}_{\rm par}$ to ``catch up'' with 
$\overline{\Delta t}_{\rm ser}$. The drawback is that  
employing large blocks can lead to excessive memory usage.
The PE update rate of the $n$-fold way routine with $l$$=$$1024$ slightly 
decreases compared to $l$$=$$512$ (Fig.\ \ref{fixed_pe}c), 
even though the utilization and the mean
time increment slightly increase. In fact, $l$$=$$1024$ is fairly close to 
the largest block size we could allocate in the memory of one node.
For very small $l$ ($l$$\leq$$16$), the utilization and $\overline{\Delta t}$ 
of the $n$-fold way routine are so low that the routine is actually 
outperformed by the parallel Metropolis one (Fig.\ \ref{fixed_pe}c). 

\subsection{Simulation with fixed number of PEs and varying physical control 
parameters.} 
\begin{figure}[t]
\begin{center}
\epsfxsize=7.0cm\epsfysize=7.0cm\epsfbox{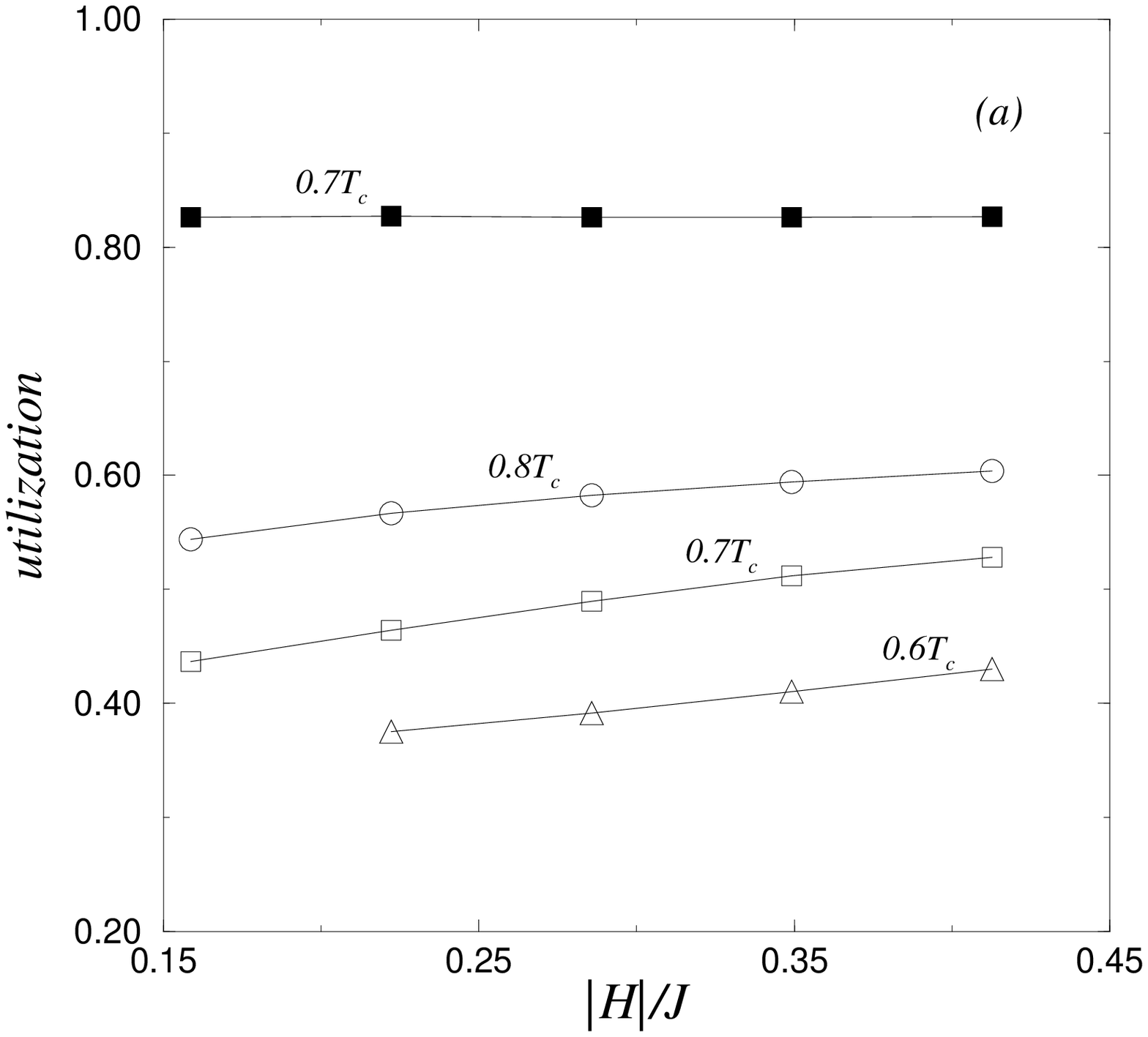} \hspace*{1cm}
\epsfxsize=7.0cm\epsfysize=7.0cm\epsfbox{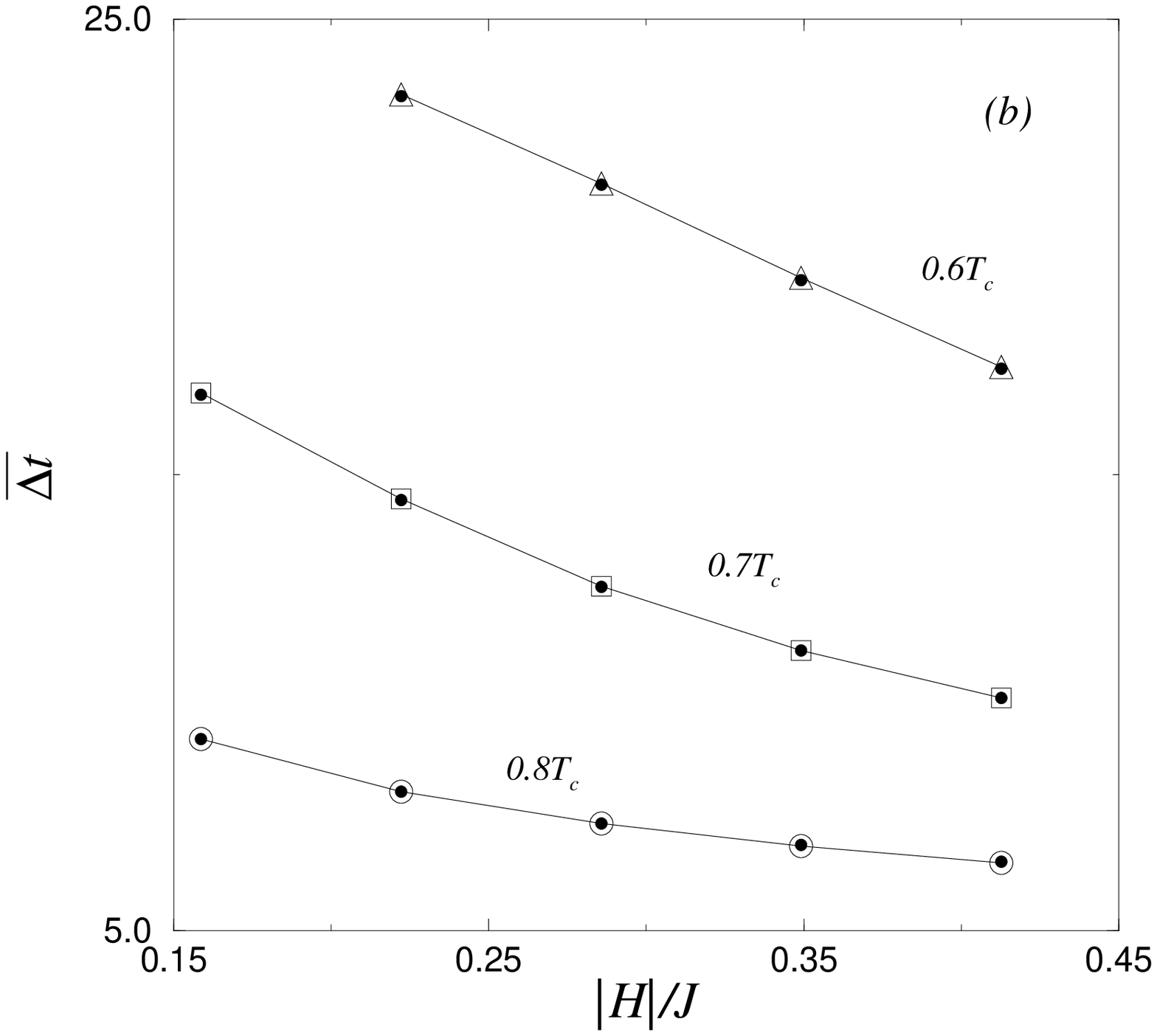} \\ \vspace*{1cm}
\epsfxsize=7.0cm\epsfysize=7.0cm\epsfbox{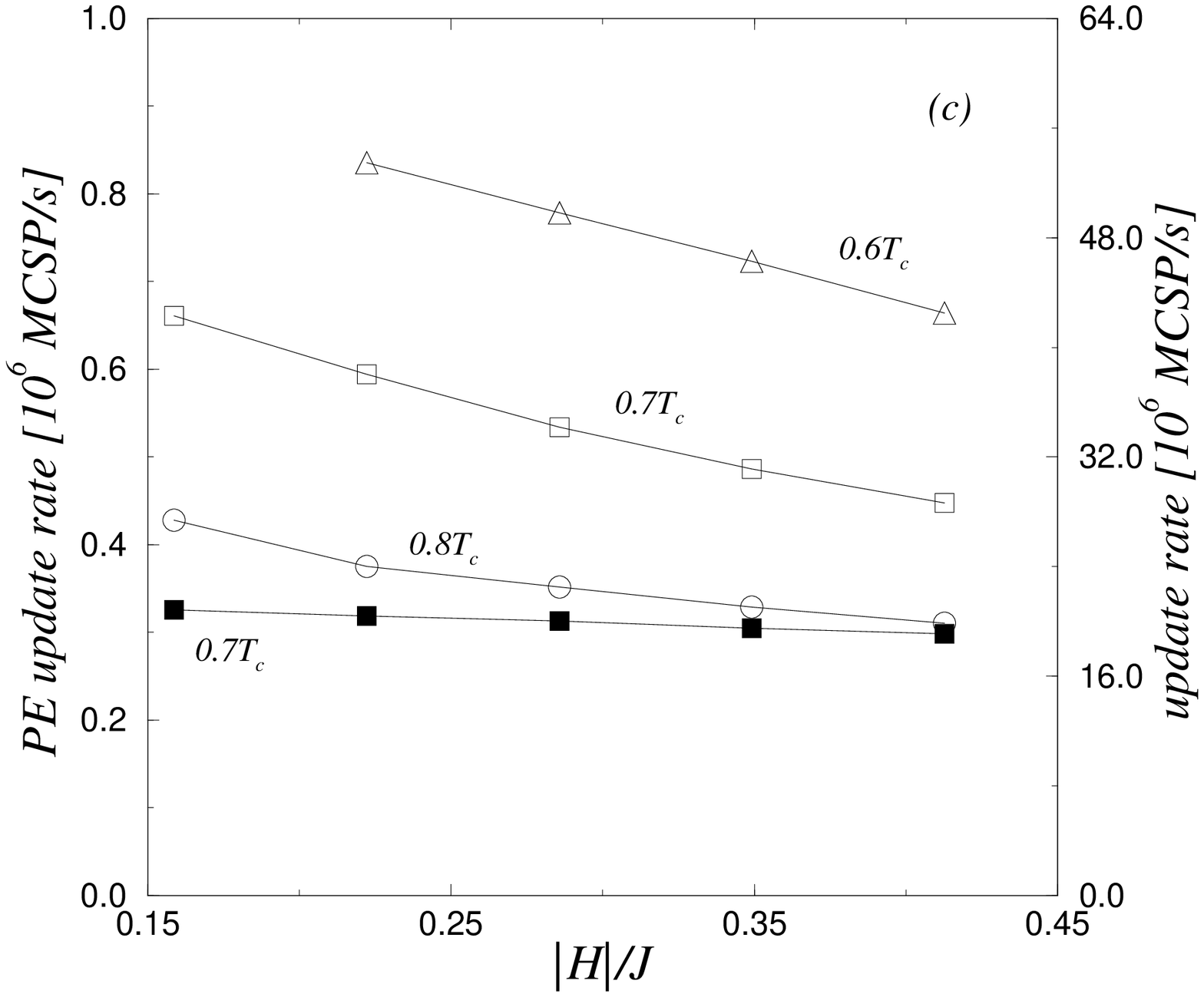} \hspace*{1cm}
\epsfxsize=7.0cm\epsfysize=7.0cm\epsfbox{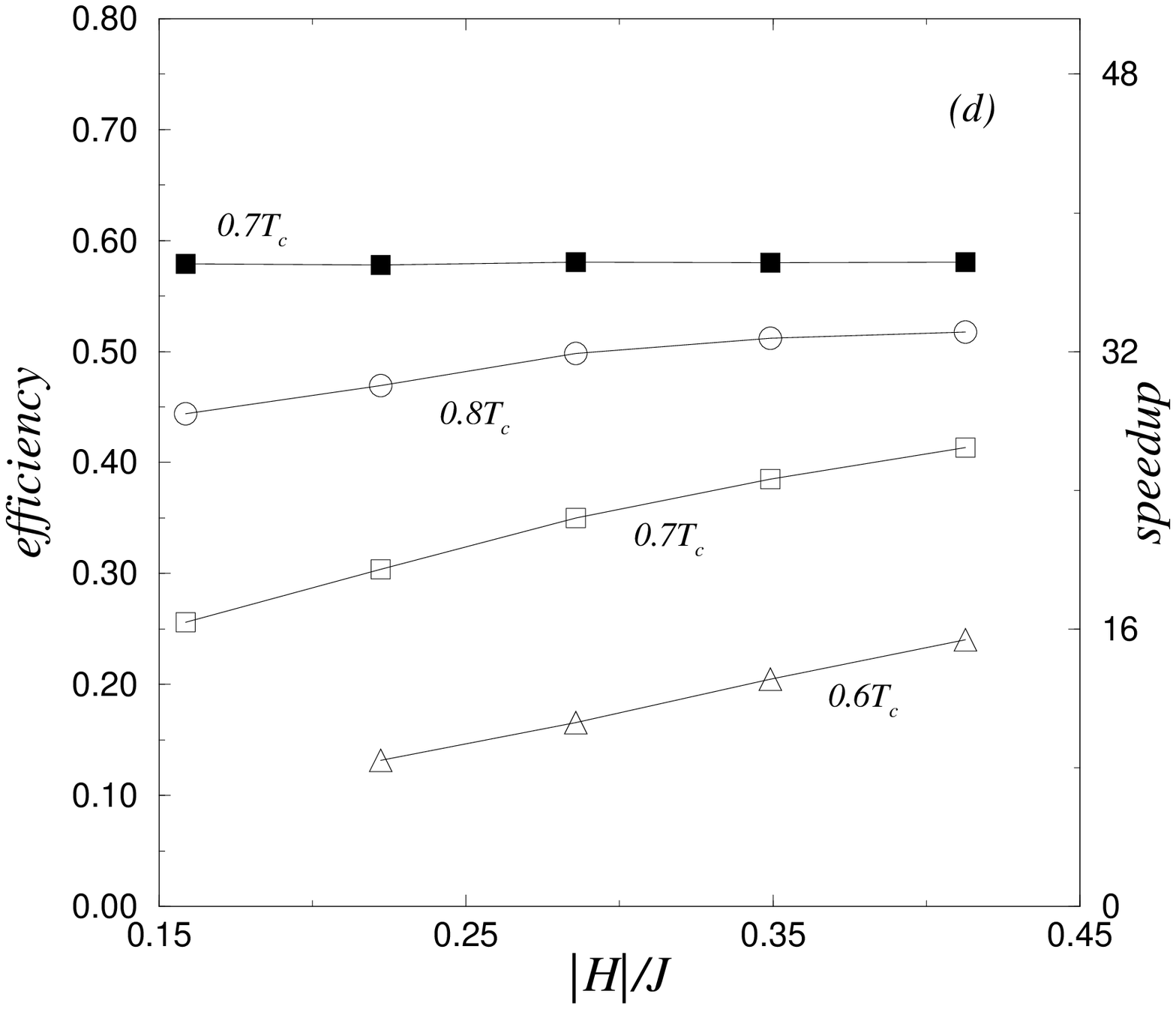}
\end{center}
\caption{Performance analysis for three different temperatures, 
$T$$=$$0.8T_{c}, 0.7T_{c}$, and $0.6T_{c}$ for the parallel $n$-fold way 
algorithm (open circles, squares, and triangles, respectively), and 
$T$$=$$0.7T_{c}$ for the parallel Metropolis algorithm (filled squares) 
as functions of the magnetic field. We employ $N_{\rm PE}$$=$$64$ and $l=128$ 
($L$$=$$1024)$. The lines connecting the data points are merely guides to 
the eye. 
(a) Utilization. 
(b) Mean time increment, $\overline{\Delta t}$, for the parallel $n$-fold way 
algorithm. Filled circles indicate the theoretical predictions of Eq.\
(\protect\ref{inc_degrade}).
(c) Update rate (right scale) and PE update rate (left scale).
(d) Speedup (right scale) and efficiency (left scale).}
\label{var_TH}
\end{figure}

Here $N_{\rm PE}$$=$$64$ and $l$$=$$128$ are kept constant while 
we vary the magnetic field for three different temperatures to study the 
effects of changing the characteristic length and time scales of the simulated
system. 
As expected, the utilization for the Metropolis algorithm is
{\em unaffected} (Fig.\ \ref{var_TH}a); the slight increase in the PE update 
rate at lower fields (Fig.\ \ref{var_TH}c) or temperatures is due to the 
fact that for higher rejection rates fewer variables need to be updated.
The parallel $n$-fold way routine suffers from low utilization at low 
temperatures and fields: in a high percentage of the execution time of the 
main simulation cycle, a spin on the boundary is picked. This not only implies
necessary communications with its neighboring PE(s), but possible
idling if the local times of the neighboring blocks are behind.
Although the time increments increase with decreasing 
temperature and field, they cannot exceed $l/4$ as shown in 
\mbox{Eq.\ (\ref{inc_par})} (Fig.\ \ref{var_TH}b). Nevertheless, even with the
bounded time increments, the parallel $n$-fold way routine outperforms the
parallel Metropolis one (Fig.\ \ref{var_TH}c). The drawback of the parallel
(partially rejection-free) $n$-fold way scheme is the low efficiency with 
respect to its {\em serial} counterpart (Fig.\ \ref{var_TH}d); for example at 
$T$$=$$0.6T_{c}$ and $|H|/J$$=$$0.2222$ it is only $13.2\%$, corresponding to 
a speedup of only $8.4$. For comparison, the mean time increments of the 
serial $n$-fold way routine are also given in Table \ref{table2}.

\section{Applications}
As discussed in Section II, below the equilibrium critical temperature 
the kinetic Ising system exhibits metastable decay after an instantaneous 
magnetic field reversal from $|H|$ to $-|H|$. Using standard
droplet theory \cite{switch}, one can show that a thermally nucleated 
domain of the stable phase must reach a critical droplet size, corresponding 
to a (temperature and field dependent) critical radius $R_c$, 
before its growth becomes energetically favorable.
For systems significantly larger than the typical droplet separation $R_o$ 
(which decreases in a nonlinear fashion with increasing $|H|$ and $T$), many 
droplets of the stable phase form and grow until they coalesce and occupy the 
whole system (multi-droplet (MD) regime) \cite{switch,Rafael}. The parameters 
we choose correspond to this decay mode. Note that $R_c$$\ll$$R_o$, in 
particular, $R_{c}/R_{o}\rightarrow 0$ in the $|H|\rightarrow 0$ limit.
A quantity of interest is the lifetime of the metastable phase, 
$\langle\tau\rangle$, which is 
defined as the average first-passage time to zero magnetization. 
Exploiting that the system is self-averaging in this regime, one may employ
the classical Avrami's law for homogeneous systems \cite{Avrami} in two
dimensions to obtain an analytical form for the time evolution of the system 
magnetization \cite{switch,Rafael}:
\begin{equation}
m(t)\approx (m_{\rm ms}-m_{\rm s})\phi_{\rm ms}(t) + 
m_{\rm s}\;,
\label{avrami}
\end{equation}
where
\begin{equation}
\phi_{\rm ms}(t)=e^{ -(\ln2) (t/\langle\tau\rangle)^{3} }
\label{meta_volume}
\end{equation}
is the volume fraction of the metastable phase, 
and $m_{\rm ms}$ and  $m_{\rm s}$ are the metastable and the stable 
(equilibrium) magnetization, respectively.

\begin{figure}[t]
\begin{center}
\epsfxsize=6.0cm\epsfysize=6.0cm\epsfbox{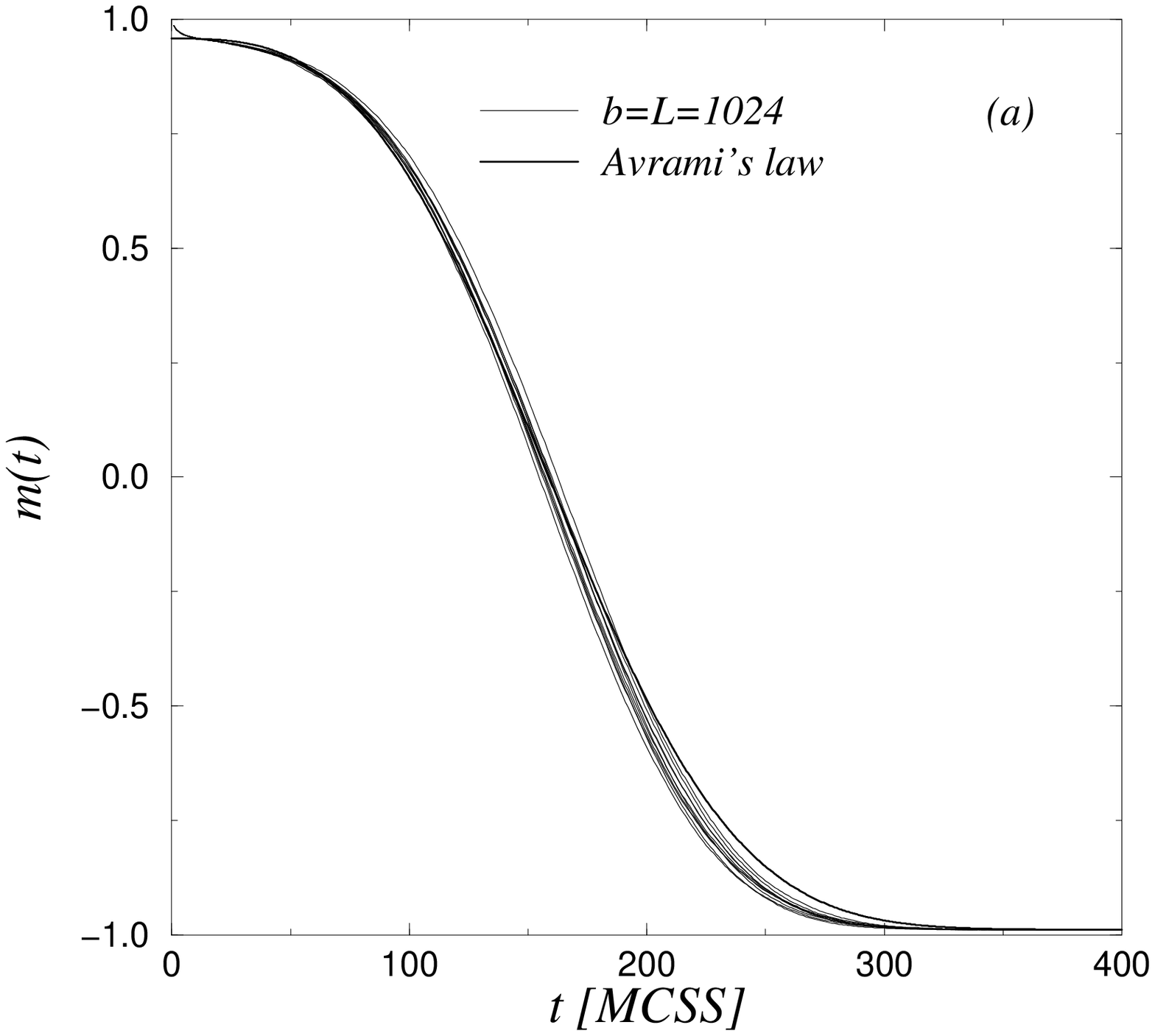} \hspace*{1cm}
\epsfxsize=6.0cm\epsfysize=6.0cm\epsfbox{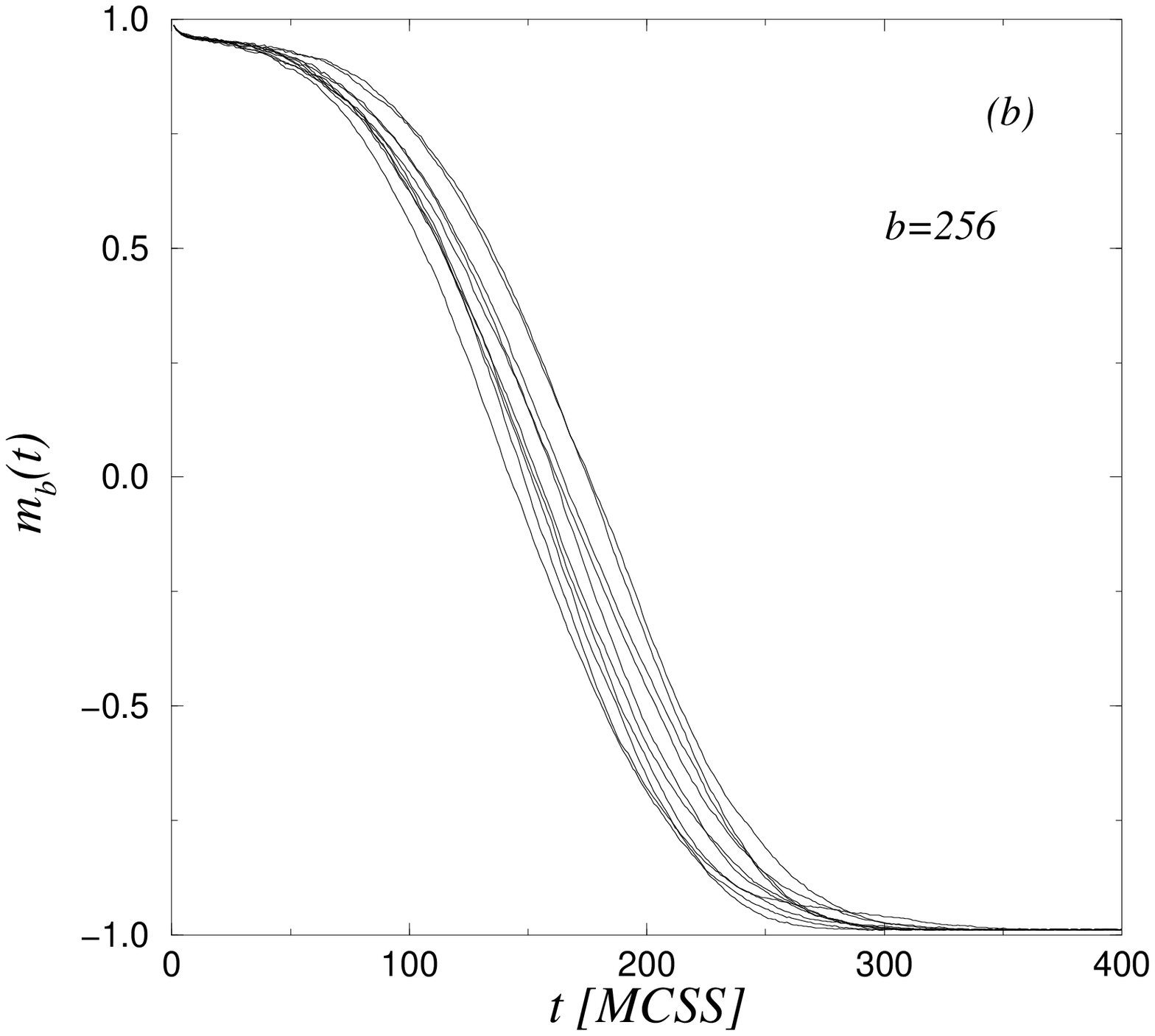} \\ \vspace*{1cm}
\epsfxsize=6.0cm\epsfysize=6.0cm\epsfbox{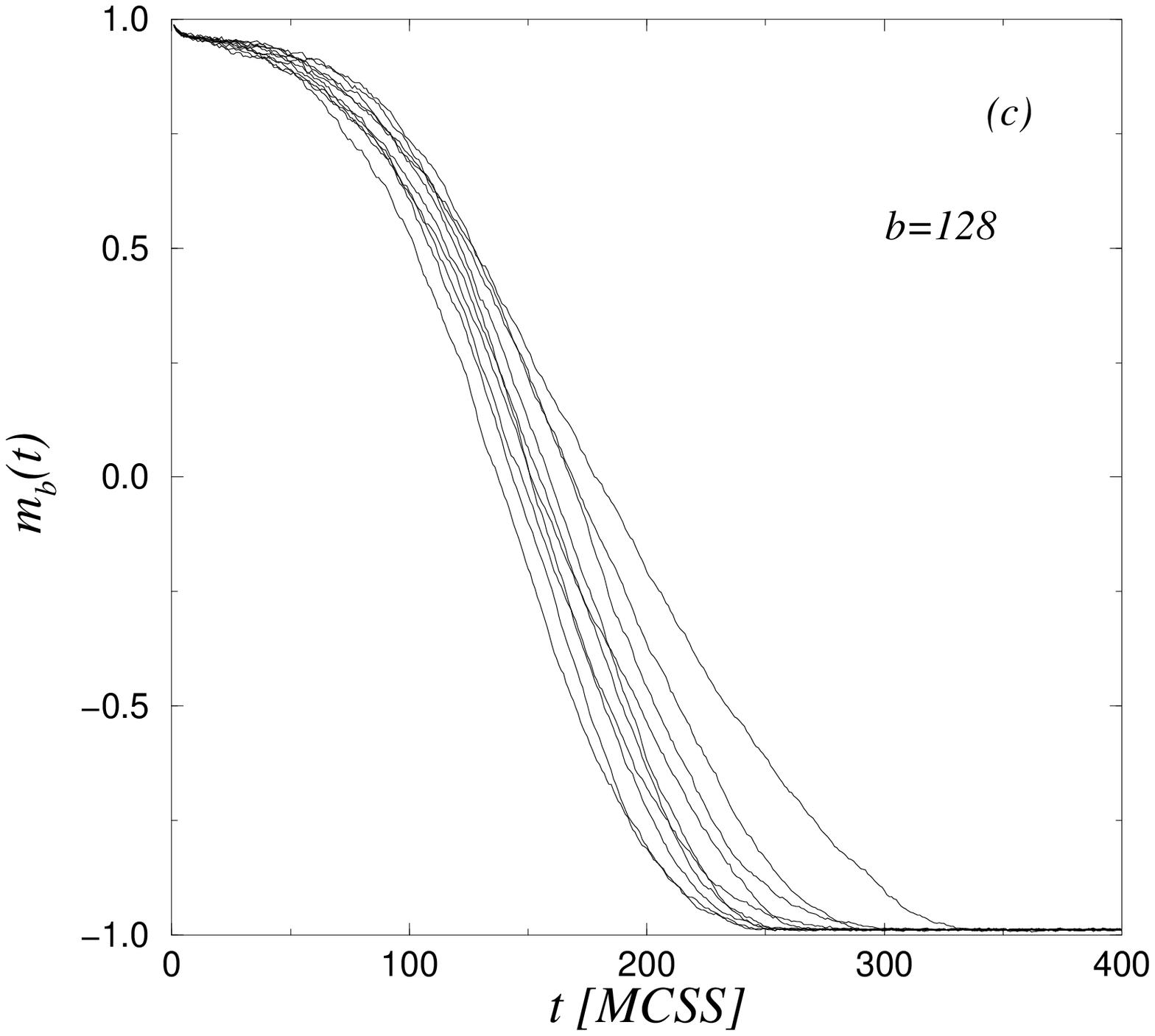} \hspace*{1cm}
\epsfxsize=6.0cm\epsfysize=6.0cm\epsfbox{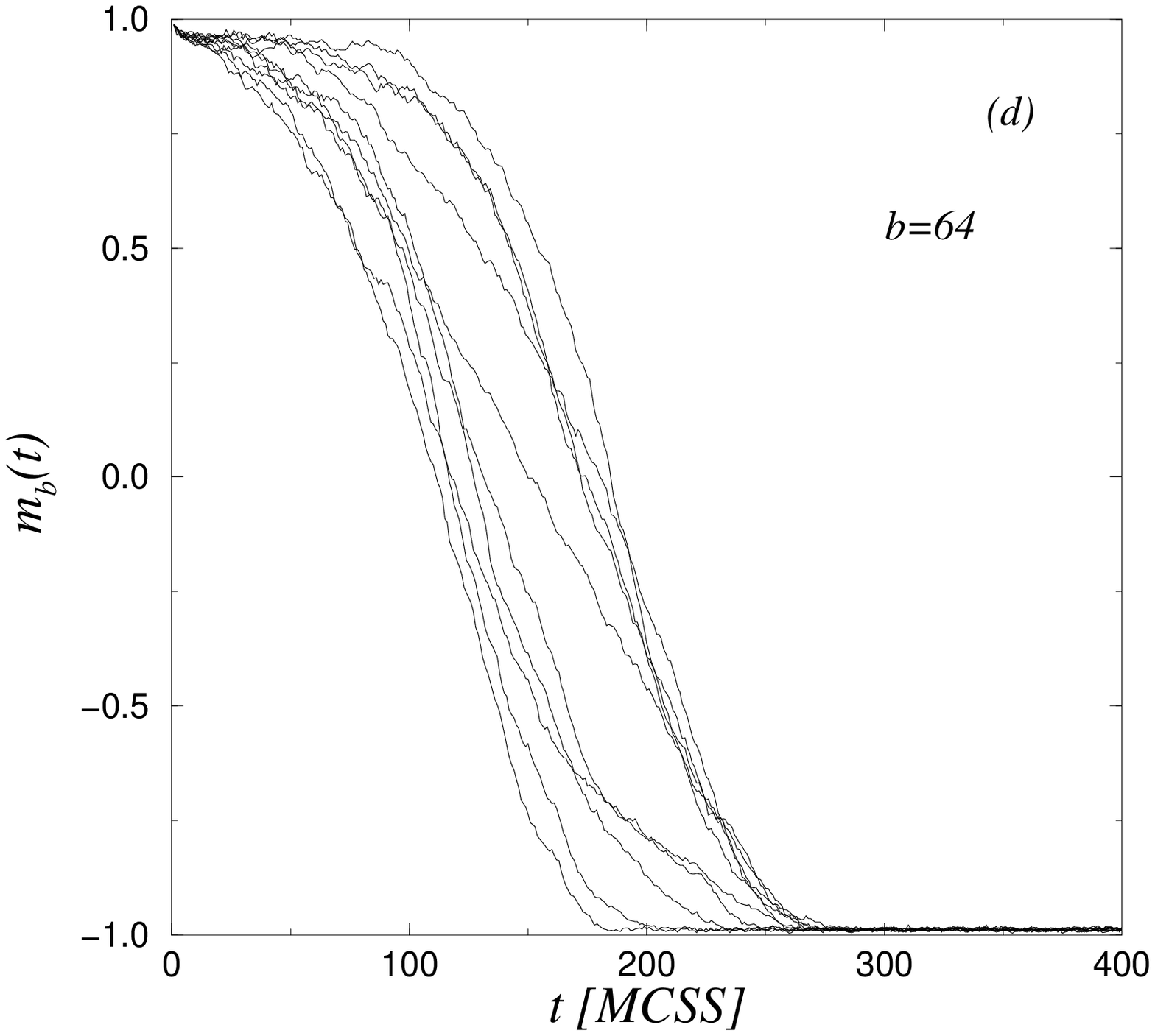} \\ \vspace*{1cm}
\epsfxsize=6.0cm\epsfysize=6.0cm\epsfbox{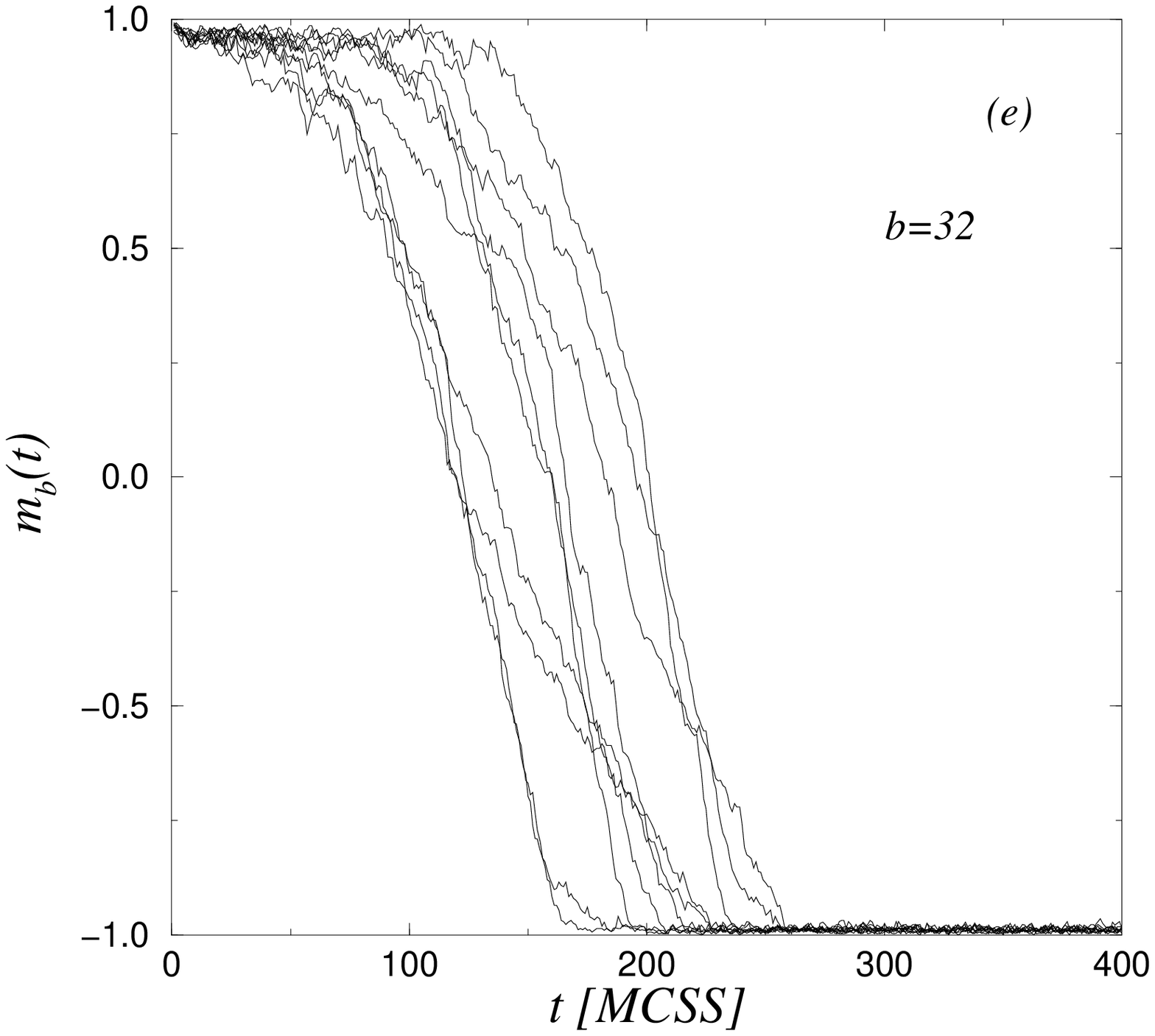} \hspace*{1cm}
\epsfxsize=6.0cm\epsfysize=6.0cm\epsfbox{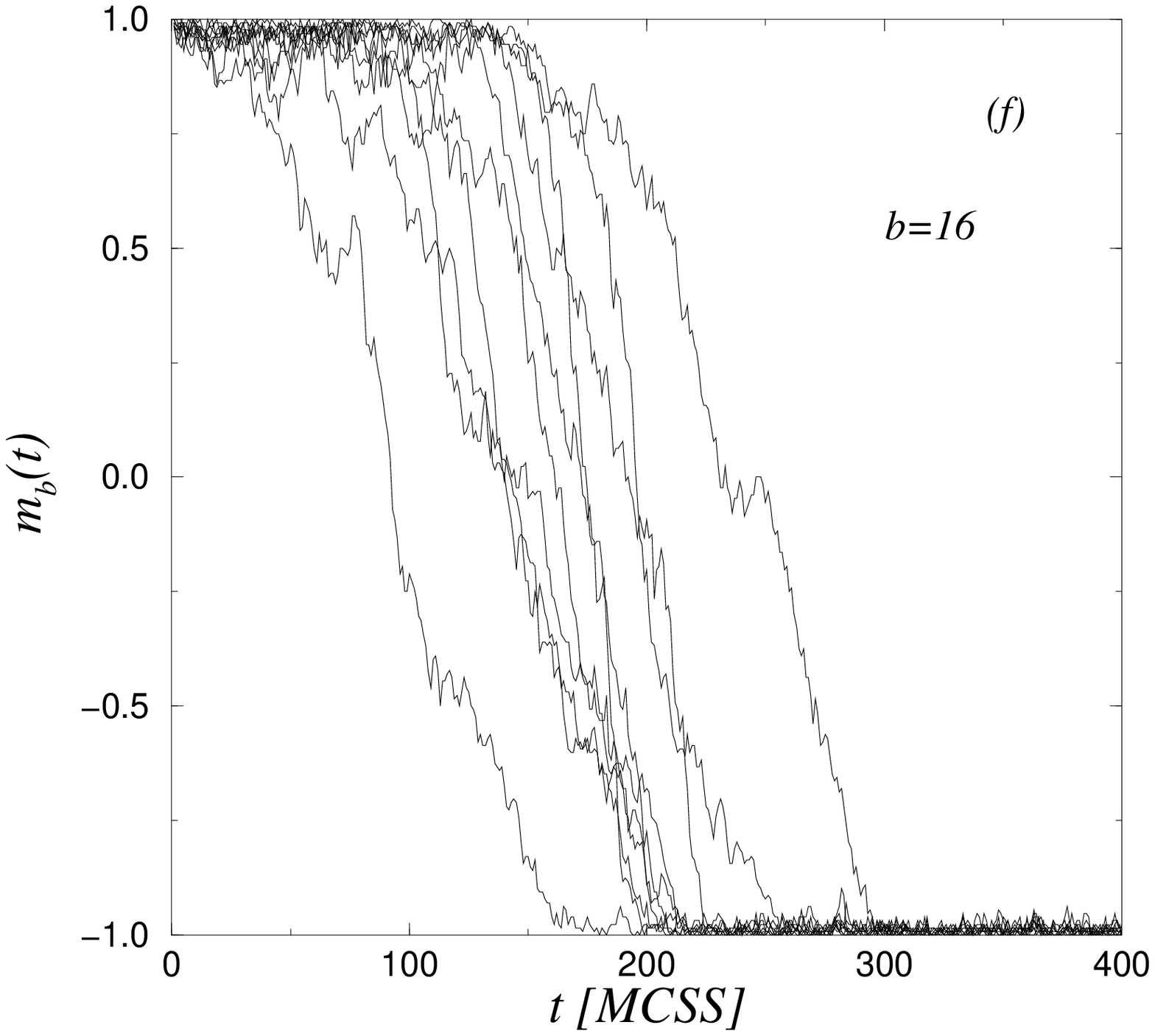}
\end{center}
\caption{Global and block magnetization time series for 10 different 
realizations at $T$$=$$0.7T_{c}$ and $|H|/J$$=$$0.2857$.
(a) For $b$$=$$L$$=$$1024$ (global). The thick solid line, which is difficult 
to distinguish from the simulation results, represents Avrami's law 
[Eqs. (\protect\ref{avrami}) and (\protect\ref{meta_volume})].
(b) For $b$$=$$256$.
(c) For $b$$=$$128$.
(d) For $b$$=$$64$.
(e) For $b$$=$$32$.
(f) For $b$$=$$16$.}
\label{magn_series}
\end{figure}

\begin{figure}[t]
\begin{center}
\epsfxsize=7.0cm\epsfysize=7.0cm\epsfbox{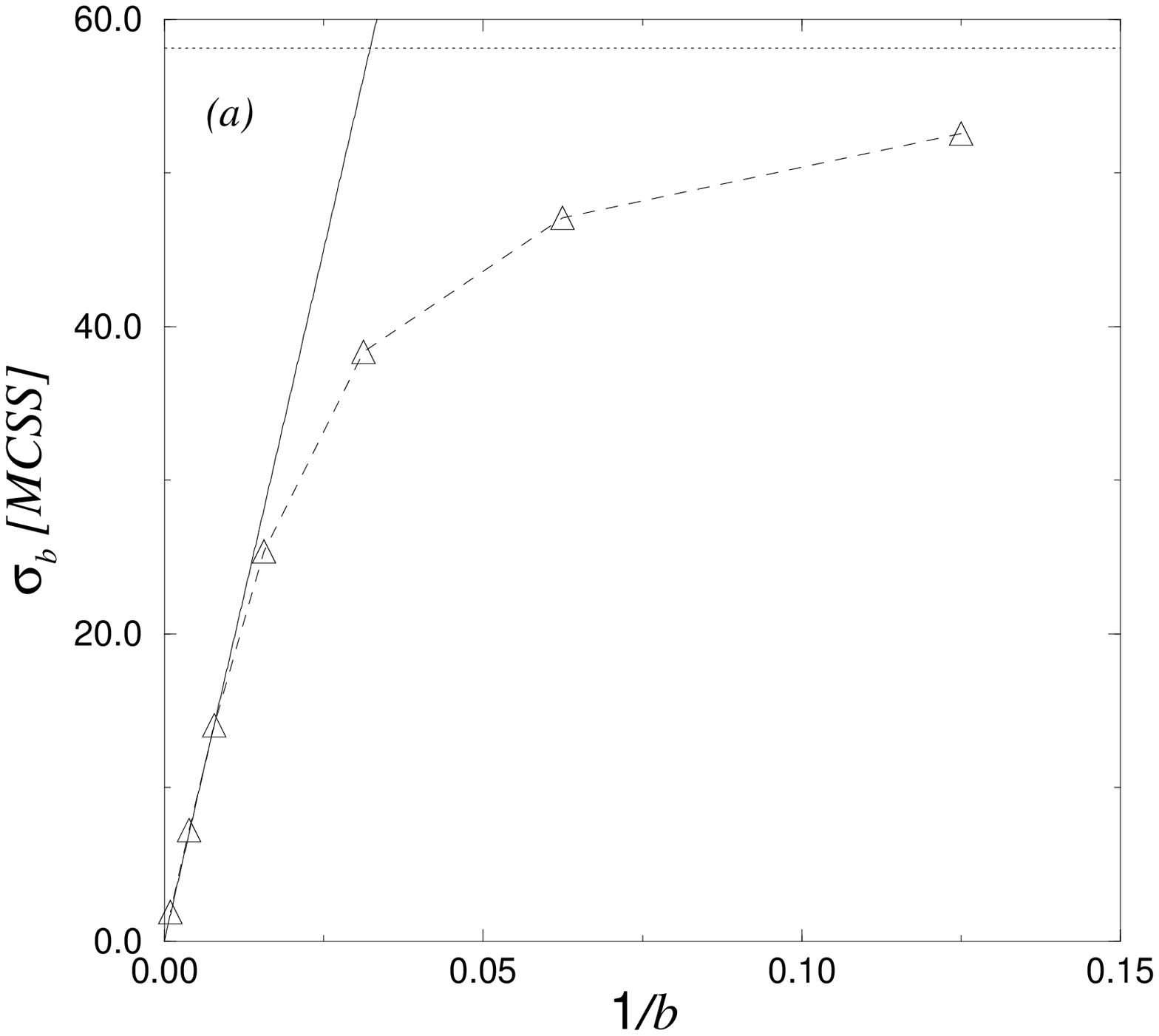} \hspace*{1cm}
\epsfxsize=7.0cm\epsfysize=7.0cm\epsfbox{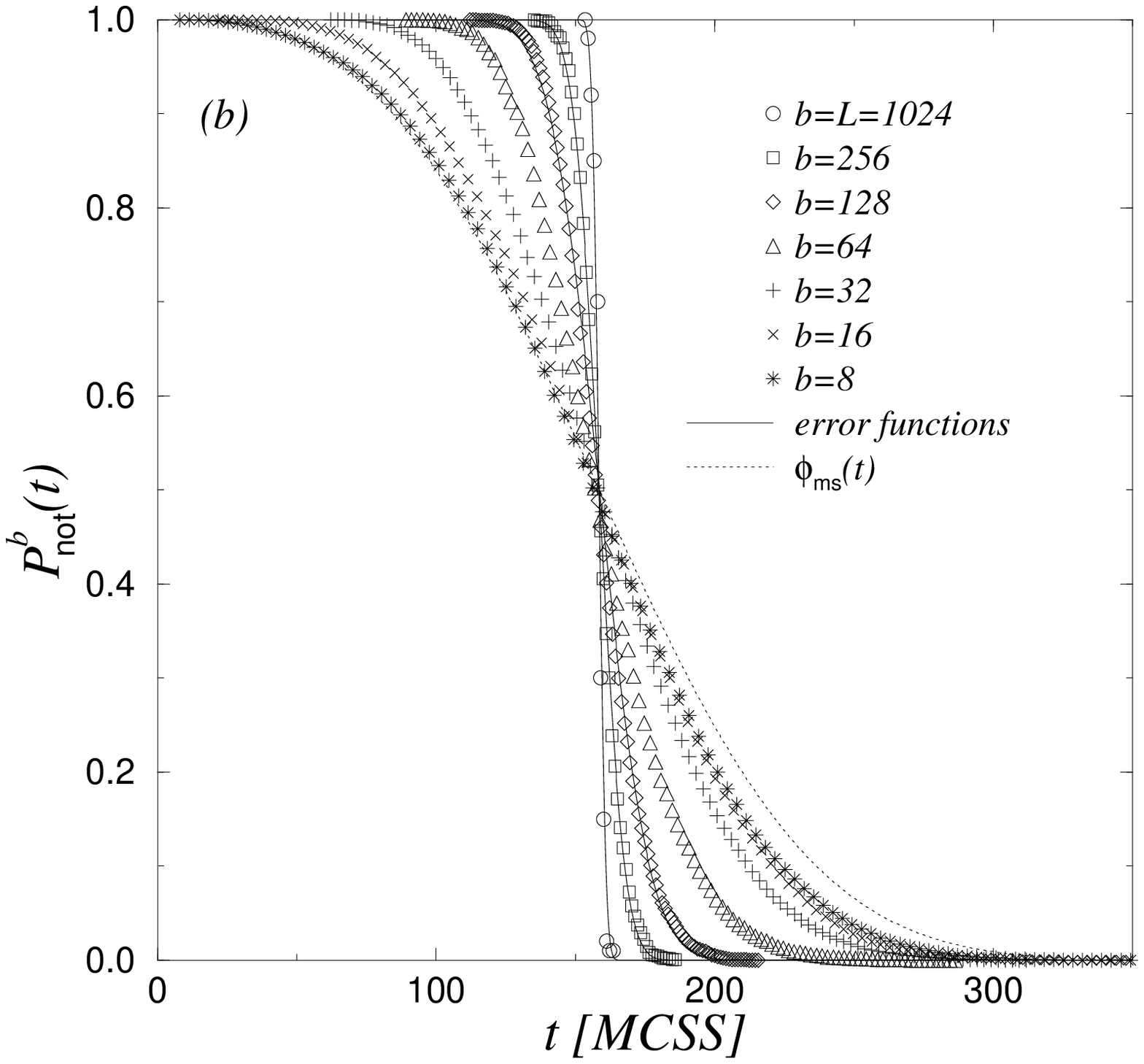}
\end{center}
\caption{(a) Standard deviation of the first-passage time of the 
block magnetization to zero (open triangles) as a function of the inverse
block size, $1/b$. 
The dashed line is merely a  guide to the eye.
The solid straight line represents the theoretical dependence on $b$ of the 
standard deviation for large $b$. The dotted horizontal line is the 
$b/R_{o}\rightarrow 0$ limit of the standard deviation 
($\sigma_{o}=58.12$ MCSS) within the coarse-grained approximation.
(b) The probability that the block magnetization, observed 
through a $b$$\times$$b$ window, has not changed sign by time $t$ after
a sudden magnetic field  reversal in a $1024$$\times$$1024$ Ising system.
The solid curves for $b$$=$$1024,256,\;\mbox{and}\; 128$ are  scaled error 
functions, given by theory \protect\cite{switch} for large $b$. The dotted 
curve for $b$$=$$8$ represents the theoretical coarse-grained 
$b/R_{o}\rightarrow 0$ limit, i.e., the metastable volume fraction, 
$\phi_{\rm ms}(t)$ [Eq. (\protect\ref{meta_volume})], 
given by Avrami's law. As expected, it fits the data very well for 
$t$$<$$\langle\tau\rangle$, where droplet coalescence is unimportant.}
\label{P_not}
\end{figure}

For illustration we choose an $L$$=$$1024$ system at
$T$$=$$0.7T_{c}$ and $|H|/J$$=$$0.2857$, and we study metastable
decay as observed through a $b$$\times$$b$ window. Employing different 
block sizes, $b$, mimics the effect of choosing 
different finite (smaller than the system size) observation windows in a 
large system. This is clearly relevant to real experiments and observations,
such as those using the magnetooptical Kerr effect \cite{Kerr}. 
We run our application with $N_{\rm PE}$$=$$16, 64$, and $256$, directly 
corresponding to block sizes $b$$=$$l$$=$$256, 128$, and $64$, respectively. 
Since the maximum number of
PEs available to us is $512$, for $b$$\leq$$32$ (for the same $L$$=$$1024$ 
system) we employ $256$ PEs with $l$$=$$64$, and monitor observables for a 
$b$$\times$$b$ block within the $64$$\times$$64$ subsystem carried by each PE.
Even our smallest block, $b$$=$$8$, is sufficiently large that subcritical
fluctuations of size $R<R_{c}\approx 2$ are not recorded as first-passage 
times to zero block magnetization.
Together with the global magnetization, Eq.\ (\ref{magn}), we monitor the 
individual block magnetizations
\begin{equation}
m_{b}=\frac{1}{b^2} \sum_{i=1}^{b^2} s_i\;.
\end{equation}
Typical time series of these quantities are shown in Fig.\ \ref{magn_series},
where time is measured in units of MCS/spin (MCSS).
Recording the first-passage times to zero for at least $100$ escapes, we
calculate the mean and the standard deviation of the lifetime 
(Fig.\ \ref{P_not}a) and construct histograms for $P^{b}_{\rm not}(t)$, the 
probability that the system or block magnetization has not changed sign by 
time $t$ (Fig.\ \ref{P_not}b). 

The {\em average} lifetime of a finite subsystem, as observed through a finite 
window, {\em does not} differ significantly from the lifetime of the whole 
system ($\langle\tau\rangle$$=$$158.5$ MCSS). However, the statistical 
properties of the lifetime change not only quantitatively, but also 
qualitatively with $b$. For $b$ 
much larger than the typical droplet separation ($R_{o}\approx 32$
lattice constants at this temperature and field), the time-dependent block 
magnetization itself is self-averaging, the switching process is Gaussian, and
consequently $P^{b}_{\rm not}(t)$ is an error function \cite{switch}. 
While the average lifetime within a block is unaffected, its standard 
deviation, $\sigma_b$, is proportional to $1/b$ in two dimensions. Once $b$ 
becomes comparable to or smaller than $R_o$, the 
above picture is no longer valid and a crossover to a qualitatively different 
behavior is observed. For these smaller blocks, the switching within a block 
is not related to 
the growth of several droplets nucleated within that block, but rather to 
a single droplet formed within the same block, or to droplets formed 
elsewhere in the system which propagate across the observed block. 
It is a challenging theoretical problem to describe the switching 
behavior analytically in this crossover regime, and work is in progress to
accomplish this task. In the $b/R_{o}\rightarrow 0$ limit the
coarse-grained approximation (in which the size of the critical droplet is
negligible) yields $P^{o}_{\rm not}(t)\approx\phi_{\rm ms}(t)$ 
[Eq. (\ref{meta_volume})], since the
probability that the block magnetization has not switched by time 
$t$ then becomes equal to the volume fraction of the metastable phase.
The standard deviation in this limit can be obtained from the probability
density of the first-passage-time, $-dP^{o}_{\rm not}/dt$, 
yielding $\sigma_{o}\approx 0.1345\langle\tau\rangle\approx 58.12$ MCSS.
Note, however, that for $b$ smaller than the diameter of the critical droplet
($2R_{c}\approx4$ lattice constants for our parameters) the above argument 
(which is based on the coarse-grained picture) is no longer valid, 
since zero-crossings of the block magnetization are frequently induced by 
subcritical droplets. 

Another future application in the MD regime is to study the
response to a periodic applied magnetic field, which exhibits highly 
nontrivial hysteretic behavior \cite{sides}. 
If the half-period of the applied field is less than the metastable lifetime 
$\langle\tau\rangle$, the system almost always does not  switch, resulting in 
a nonzero period-averaged magnetization (``dynamically ordered phase''). 
On the other hand, when the half-period exceeds $\langle\tau\rangle$, the 
magnetization switches in almost every half-period and the period-averaged
magnetization is zero (``dynamically disordered phase''). 
The transition between these two 
phases is sharp and singular: the system exhibits a {\em dynamic} phase
transition, which fits into the general framework of critical phenomena and
continuous phase transitions \cite{crit_phenom}.
Here there is clearly a need to study scaling and universality by obtaining 
the corresponding critical exponents with high accuracy. Also in this regime, 
our parallel algorithm appears to be very efficient for large systems. 
We plan to implement it with a periodic square-wave 
shaped applied magnetic field and carry out  a large-scale finite-size 
scaling analysis of the dynamic phase transition.

\section{Conclusion}
We have studied the performance of the parallel $n$-fold way dynamic Monte
Carlo algorithm proposed by Lubachevsky \cite{Luba}, in which each PE carries
an $l$$\times$$l$ block of random variables. The algorithm was 
implemented for a two-dimensional kinetic Ising ferromagnet undergoing
metastable decay, but the
parallel scheme is generically applicable to a wide range of stochastic
cellular automata where discrete events (updates) are Poisson arrivals.
To obtain reasonable performance on the T3E distributed-memory parallel
architecture and to be faithful to the original dynamics, one must utilize an 
asynchronous update scheme with continuous time. Then the expensive global 
barrier synchronizations are avoided and spin-flip attempts are modeled as 
independent Poisson arrivals. 
We analyzed the performance of our implementation,
which sensitively depends on the block size and the number of PEs, as well as 
on the characteristic length and time scales of the simulated system. 
We found that for large enough block size, the routine 
outperforms the standard parallel Metropolis algorithm. For moderately
low temperatures it yields high speedups 
with respect to the already fast serial $n$-fold way algorithm. For example, 
at $T$$=$$0.7T_{c}$ and $|H|/J$$=$$0.2857$, employing $l$$=$$256$ we obtained
a speedup of $260$ with $400$ PEs, and for $l$$=$$1024$
a speedup of $58$ with $64$ PEs. 
Often the system size (possibly large) is specified, and  
for fixed $L$, although significantly worse than linear, the speedup is still
a monotonically increasing function of the number of PEs, up to the maximum 
$256$ PEs that we studied. 
At the same time, the efficiency is monotonically 
decreasing, which results in larger {\em total} CPU time usage to execute the 
same task with a larger number of PEs. If one has unlimited resources
(i.e., no allocation limits) this aspect is not relevant. For most, like
us, who have limited CPU resources on a certain parallel architecture, 
``optimization'' between speedup and efficiency can be important.   
Our implementation is obviously best suited to simulating large systems.
 
On the other hand, for very low temperatures, the algorithm does not provide an
efficient way to simulate metastable decay.
The reason for the relatively narrow regime of efficient implementation 
lies in the introduction of a special class in the $n$-fold way 
algorithm which ``shields'' the blocks from each other, but
significantly decreases the typical time increments. 
The algorithm avoids rollbacks, but pays a large price: 
it looses the arbitrarily large time increments that is the
most important feature of the serial $n$-fold way algorithm, at 
arbitrarily low temperature and field. To obtain reasonable efficiency compared
to the serial $n$-fold way algorithm,
one needs to employ large blocks such that 
$l/4 \stackrel{>}{\sim} \overline{\Delta t}_{\rm ser}$, and clearly it is 
impossible to keep up with very large serial time increments by increasing $l$.

One way to preserve the advantage of the original $n$-fold way algorithm
in principle would be to apply it directly on each block (optimistic approach).
This would require a complex protocol to correct erroneous computations.
Such a rollback procedure would ensure the correct time ordering of simulated 
events. 
This mechanism is not unknown in distributed event simulation \cite{rollback}
and certainly has potential in our model. The complexity of such an 
implementation, however, might carry a tremendous overhead with respect to the 
very simple and fast serial algorithm for the Ising model.
Another possible way to improve efficiency, while avoiding
a general rollback procedure, is to consider relaxation 
\cite{relaxation} which might use local speculative computations 
before scheduling an event \cite{private}.

\section*{Acknowledgments}
Special thanks to B. D. Lubachevsky and M. Kolesik for invaluable discussions 
and to R. Gerber of the National Energy Research Scientific Computing Center
consulting group for important hints on 
debugging the parallel code.  Helpful discussions with S. W. Sides, S. J. 
Mitchell, and G. Brown are also gratefully acknowledged. This research was 
supported in part by the Florida State University Supercomputer Computations
Research Institute under US Department of Energy Contract No.\ 
DE-FC05-85ER25000, by NSF Grants No.\ DMR-9520325 and DMR-9871455, and by
the Florida State University Center for Materials Research and Technology. 
This research used resources of NERSC, which is supported by the Office of 
Energy Research of the US DOE under Contract No.\ DE-AC03-76SF00098.


\newpage

\begin{table}
\begin{tabular}{rrrrrrrr|r}
 & \multicolumn{7}{c|}{parallel $n$-fold way with block size $l$} & serial \\ 
 \hline
 $l$ & 16 & 32 & 64 & 128 & 256 & 512 & 1024 & ($\ast$) \\ 
 $\overline{\Delta t}$ & 3.7 & 6.1 & 9.2 & 12.6 & 15.4 & 17.4 & 18.5 & 19.9 \\
\end{tabular}
\caption{Mean time increments (in MCSP) for the serial and the parallel 
$n$-fold way algorithms with different block size $l$ at 
$T$$=$$0.7T_{c}$, $|H|/J$$=$$0.2857$. They are approximately independent
of the full system size $L$ and $N_{\rm PE}$.  ($\ast$) The mean 
time increment for the serial algorithm is approximately independent of $L$.}
\label{table1}
\end{table}

\begin{table}
\begin{tabular}{rrlrrrrr} 
 \multicolumn{3}{c}{\mbox{ }} & \multicolumn{5}{c}{$|H|/J$} \\
 & & & 0.1587 & 0.2222 & 0.2857 & 0.3492 & 0.4127 \\ \hline  
 & 0.6 & serial   &  - \mbox{ }   & 81.5 & 61.4 & 46.4 & 36.3 \\ 
 &     & parallel &  - \mbox{ }   & 23.4 & 21.4 & 19.3 & 17.4 \\ \cline{2-8}
$T/T_{c}$ & 0.7 & serial   & 33.8 & 25.4 & 19.9 & 16.5 & 14.3 \\ 
          &     & parallel & 16.8 & 14.5 & 12.6 & 11.1 & 10.1 \\ \cline{2-8}
 & 0.8 & serial   & 12.5 & 10.4 &  9.2 &  8.5 &  7.9 \\ 
 &     & parallel &  9.2 &  8.0 &  7.4 &  6.9 &  6.5  
\end{tabular}
\caption{Mean time increments (in MCSP) for the serial and the parallel 
$n$-fold way algorithms for different temperatures and magnetic fields
($N_{\rm PE}$$=$$64$, $l$$=$$128$).}
\label{table2}
\end{table}

\newpage

\newpage

\newpage

\newpage

\newpage

\newpage

\end{document}